\documentclass[aps,prx,reprint,superscriptaddress,amsmath,amssymb,showpacs,floatfix,longbibliography]{revtex4-1}

\usepackage{graphicx}
\usepackage{dcolumn}
\usepackage{bm}
\usepackage{color}
\usepackage[utf8]{inputenc}
\usepackage{physics}
\usepackage{mathtools}
\usepackage{soul}

\newcommand{\rT}{\textrm{\tiny T}}

\begin{document}
\title{Thermodynamic uncertainty relation in the overdamped limit with a magnetic Lorentz force}
\author{Jong-Min Park}
\email{jmpark19@kias.re.kr}
\affiliation{School of Physics, Korea Institute for Advanced Study, Seoul 02455, Korea}
\author{Hyunggyu~Park}
\email{hgpark@kias.re.kr}
\affiliation{School of Physics, Korea Institute for Advanced Study, Seoul 02455, Korea}
\affiliation{Quantum Universe Center, Korea Institute for Advanced Study, Seoul 02455, Korea}
\date{\today}

\begin{abstract}
{
In nonequilibrium systems,
the relative fluctuation of a current
has a universal trade-off relation with the entropy production,
called the thermodynamic uncertainty relation (TUR).
For systems with broken time reversal symmetry, its violation
has been reported in specific models or in the linear response regime.
Here, we derive a modified version of the TUR analytically
in the overdamped limit for general Langevin dynamics with a magnetic Lorentz
force causing  time reversal broken.
Remarkably, this modified version is simply given by the conventional TUR
scaled by the ratio of the reduced effective temperature of the overdamped motion
to the reservoir temperature, permitting a violation of the conventional TUR.
Without the Lorentz force, this ratio becomes unity and the conventional TUR is restored.
We verify our results both analytically and numerically in a specific solvable system.
}
\end{abstract}

\maketitle

\section{Introduction}\label{sec:introduction}
The thermodynamic uncertainty relation (TUR) states that
current fluctuation has a universal trade-off relation with thermodynamic
cost~\cite{Barato2015thermodynamics,Seifert2018stochastic,Seifert2019stochastic}.
To be precise, the product of the relative error square of an accumulated current
$\Phi$ and the total entropy production $\langle\Delta S^\textrm{tot}\rangle^\mathrm{ss}$ in the steady state is bounded from below as
\begin{align}\label{eq:convTUR}
{\cal Q}_\Phi \equiv \frac{\textrm{Var}^\mathrm{ss}[\Phi]}{{(\langle\Phi\rangle}^\mathrm{ss})^2} \langle\Delta S^\textrm{tot}\rangle^\mathrm{ss} \ge 2 k_\textrm{B}~,
\end{align}
where ${\cal Q}_\Phi$ is called the TUR factor for the current $\Phi$, the variance $\textrm{Var}^\mathrm{ss}[\Phi]\equiv\langle \Phi^2\rangle^\mathrm{ss} -(\langle\Phi\rangle^\mathrm{ss})^2$ with  a steady-state average $\langle \cdot\rangle^\mathrm{ss}$, and
the Boltzmann constant $k_\textrm{B}$.
As an example, consider a molecular motor in a directional stochastic motion.
Stochastic fluctuations and heat dissipation are two key quantities needed to be minimized
for ideal performance. The above TUR enlightens that both quantities can not be minimized simultaneously
and there should be a trade-off between them.

After the first discovery of the TUR for systems in the linear response regime and Markov processes on simple networks~\cite{Barato2015thermodynamics}, many studies have explored its generality and applicability
by investigating various systems~\cite{Seifert2019stochastic,Horowitz2019thermodynamic}. Exploiting a large deviation theory or an information theory, the TUR was derived for continuous-time Markov jump processes and overdamped Langevin dynamics~\cite{Gingrich2016dissipation,Gingrich2017inferring,Pietzonka2017finite,Horowitz2017proof,Hasegawa2019uncertainty,Dechant2019multidimensional}.
It has also been reported that the TUR is related
to the efficiency bound of molecular
motors~\cite{Pietzonka2016universal},
the power-efficiency trade-off relation for heat
engines~\cite{Benenti2011thermodynamic,Shiraishi2016universal,Pietzonka2018universal},
the generic stochastic equation for entropy
production~\cite{Pigolotti2017generic, chun2019universal},
the Cramer-Rao
inequality~\cite{Hasegawa2019uncertainty, Dechant2019multidimensional},
and the symmetry of the joint distribution for
the current and the entropy
production~\cite{Hasegawa2019fluctuation, potts2019thermodynamic} known as
the detailed fluctuation theorem~\cite{Jarzynski1997nonequilibrium,Crooks1999entropy,Seifert2005ep}.
Recently, various studies have developed methods for the entropy production inference based on the TUR~\cite{li2019quantifying,manikandan2020inferring,van2020entropy,otsubo2020estimating}.

Some recent studies have shown that the TUR is violated
and should be modified
when dynamics has an intrinsic time
scale~\cite{Shiraishi2017finite,Proesmans2017discrete,Barato2018bounds,Koyuk2019generalization,Barato2019unifying, koyuk2020thermodynamic},
breaks the time-reversal symmetry~\cite{Macieszczak2018unified, Brandner2018thermodynamics,Chun2019effect},
or involves any odd-parity variable under time
reversal such as velocity~\cite{Fischer2018large, Vu2019uncertainty, Lee2019thermodynamic}.
Especially, in the underdamped Langevin dynamics,
the TUR is trivially violated for reversible currents~\cite{Vu2019uncertainty} or
for finite duration time~\cite{Fischer2019free}.
Some evidence supports that the TUR could be valid for irreversible currents in the long-time limit~\cite{Fischer2019free}, but a rigorous proof is missing.
Although a variant of the TUR including dynamic activity was reported for the underdamped Langevin systems~\cite{Vu2019uncertainty} and then extended to systems with velocity-dependent forces~\cite{Lee2019thermodynamic}, their use is limited  because their  lower bounds depend on dynamic details and become trivial in the overdamped limit.

Recently, Chun {\it et al}~\cite{Chun2019effect} showed that,
in an exactly solvable underdamped Langevin system with
a magnetic Lorentz force,
work and heat currents violate the conventional TUR of Eq.~\eqref{eq:convTUR}
even in the small-inertia (overdamped) limit. This violation
may not be surprising as the Lorentz force breaks the time reversal symmetry.
Instead, they reported a modified bound for the TUR factor
in the overdamped limit for infinitely long duration time.
Interestingly, this modified lower bound is very similar to the conventional TUR bound except for a simple
dimensionless multiplication factor.
However, this system is  a special harmonic system, thus
the applicability of their finding to general systems should not be taken for granted.

In this paper, we consider a general underdamped dynamics with a magnetic Lorentz force
and rigorously show that Chun  {\it et al}'s finding is surprisingly still intact for general systems in the overdamped
limit. More remarkably, this modified TUR is valid for general currents (odd under time reversal) including
work and heat currents and even for a finite duration time. This modified TUR reads
\begin{align}
{\cal Q}_\Phi \ge 2 k_\textrm{B}\frac{T_B}{T}~,
\end{align}
 where  $T$ is the reservoir temperature and $T_B$ $(<T)$ is the {\em reduced} effective temperature of the overdamped motion, which will be defined in the next section. We note that $T_B=T$ without a Lorentz force and the conventional TUR is restored.

The derivation is rather tricky, mainly because the overdamped (small-inertia) limit  in the presence of
a Lorentz force generates a {\em non-white} noise~\cite{Chun2018emergence}. We introduce an alternative
but equivalent description with a white noise by employing a standard small-inertia expansion~\cite{Risken:1996FokkerPlanck}
on an {\em extended} Fokker-Planck equation with a current as an additional variable.
The main strategy to derive the modified TUR is based on the Cramer-Rao inequality with
a slightly different perturbation from the conventional one adopted in Refs.~\cite{Hasegawa2019uncertainty, Dechant2019multidimensional}.
We also generalize the modified TUR for an arbitrary initial state.
Finally, we calculate analytically the finite-time TUR factors for work, heat, and entropy production currents in the overdamped limit
for the solvable harmonic system with a Lorentz force in two dimensions, which confirm our modified TUR.

\section{The overdamped Langevin equation
with a magnetic field}\label{sec:II}

\begin{figure}
\includegraphics*[width=\columnwidth]{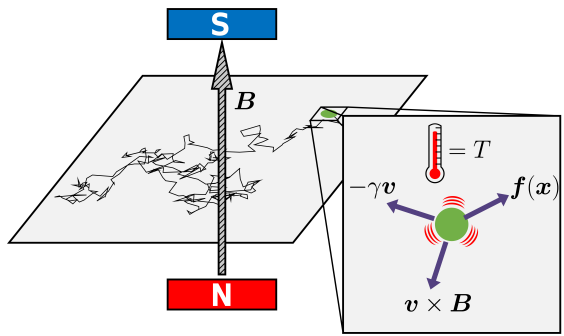}
\caption{
A illustration describing the system.
}
\label{fig:model}
\end{figure}

For convenience, we consider two-dimensional dynamics of a charged Brownian particle
in thermal contact with a heat bath at temperature $T$
and subjected to
a static magnetic field perpendicular to the motion plane.
We also consider a general in-plane force
$\boldsymbol{f}(\boldsymbol{x})$ acting on the particle, as shown in Fig.~\ref{fig:model}.
As the Lorentz force induced by the magnetic field has an in-plane component only,
the particle motion is constrained on the two-dimensional plane with a proper
initial condition.

The equation of motion of the particle is given by the
Langevin equation as
\begin{equation}\label{eq:U_Langevin}
\begin{split}
    \dot{\boldsymbol{x}} (t) & = \boldsymbol{v} (t), \\
    m \dot{\boldsymbol{v}} (t) & = \boldsymbol{f} \qty(\boldsymbol{x} (t))
    - \mathbf{G} \boldsymbol{v} (t) + \boldsymbol{\xi}_T (t)~,
\end{split}
\end{equation}
where $\dot{a}$ represents the derivative of a variable $a$ with respect to time $t$,
$\boldsymbol{x} (t)$ and $\boldsymbol{v} (t)$ are the two-dimensional position and
velocity vector of the particle, $m$ is the particle mass,
\begin{equation}\label{eq:Gamma}
    \mathbf{G} =
    \begin{pmatrix}
        \gamma & - B\\
        B & \gamma
    \end{pmatrix}
\end{equation}
is an asymmetric friction coefficient tensor
with the friction constant $\gamma$ and
the magnetic field magnitude $B$,
and $\boldsymbol{\xi}_T (t)$ is the thermal noise described by
a Gaussian white noise.
The fluctuation-dissipation relation imposes
$\expval*{\boldsymbol{\xi}_T (t) \boldsymbol{\xi}_T^\rT (t')} =
2 \gamma T \mathbf{I} \delta(t-t')$
with the identity matrix $\mathbf{I}$,
where the superscript $\vphantom{\xi}^\rT$  denotes the transpose.
The particle charge and the Boltzmann constant
are set to be unity.

%%  Fig. 1  %%%%%%%%%%%%%%%%%%%%%%%%%%%%%%%%%%%%%%%%%%
%%%%%%%%%%%%%%%%%%%%%%%%%%%%%%%%%%%%%%%%%%%%%%%%%%%%%%

Under usual experimental conditions,
the friction coefficient is so large that the inertia effects
can be ignored. The equation of motion in this overdamped (small-mass) limit
was obtained by integrating out Eq.~\eqref{eq:U_Langevin}
(see Appendix~\ref{sec:app_A} and \cite{Chun2018emergence}) as
\begin{equation}\label{eq:O_Langevin}
    \dot{\boldsymbol{x}}(t) =
    \mathbf{G}^{-1} \boldsymbol{f} \qty(\boldsymbol{x} (t)) +
    \boldsymbol{\eta}_T (t)~,
\end{equation}
where $\boldsymbol{\eta}_T (t)$ is a non-white Gaussian noise
characterized by
$\expval{\boldsymbol{\eta}_T(t) \boldsymbol{\eta}_T^\rT(t')} =
\mathbf{Z} (t-t') $
with
\begin{equation}\label{prop:co_mat}
    \mathbf{Z} (u) =
    \left \{
    \begin{matrix}
        2 T \mathbf{G}^{-1} \delta(u), & \mathrm{for}~u > 0, \\
        2 T \qty(\mathbf{G}^{-1})^\rT \delta(u), & \mathrm{for}~u < 0~,
    \end{matrix}
    \right .
\end{equation}
which is clearly {\em singular} at $u=0$.

The symmetric part of the correlation matrix $\mathbf{Z}$ is
\begin{align}
\label{eq:Zs}
&\mathbf{Z}_\mathrm{s}(u)\equiv    \frac{\mathbf{Z}(u) + \mathbf{Z}^\rT(u)}{2}
    = \frac{2 T_B}{\gamma} \mathbf{I} \delta(u) \\
&\textrm{with}\quad
    T_B = \frac{1}{1+(B/\gamma)^2} T \label{eq:TB}~,
\end{align}
indicating that the particle in the overdamped limit experiences a heat reservoir
with the effective temperature $T_B$ lower than the original reservoir temperature $T$.
This is consistent with the diffusion coefficient decrease for
a charged Brownian particle in the presence of magnetic field, reported in \cite{Czopnik2001brownian}.

The antisymmetric part of the correlation matrix plays a crucial role
for featuring nonequilibrium-ness in this subtle overdamped limit, which generates a
rotational probability current (curl flux), verified recently by numerical simulations
for various systems~\cite{Vuijk2019anomalous, Abdoli2020nondiffusive, Liao2020rectification, Vuijk2020lorentz}.
This can be more transparent by deriving the probability current directly in the Fokker-Planck description.

The systematic way to obtain a small-mass expansion of the underdamped Fokker-Planck (Kramer) equation
is well established~\cite{Risken:1996FokkerPlanck, Chun2018emergence, Durang2015overdamped}. Its
leading order gives the overdamped Fokker-Planck (FP) equation governing the time evolution of the position distribution function $p_\eta (\boldsymbol{x},t)$ as~\cite{Chun2018emergence}
\begin{equation}\label{temp:o_FP_eq}
    \partial_t p_\eta (\boldsymbol{x},t)
    = - \boldsymbol{\nabla}^\rT \cdot
    \boldsymbol{J}_\eta(\boldsymbol{x},t)~,
\end{equation}
with the probability current
\begin{equation}\label{temp:o_pro_current}
    \boldsymbol{J}_\eta(\boldsymbol{x},t) = \mathbf{G}^{-1}
    \qty( \boldsymbol{f}(\boldsymbol{x}) - T \boldsymbol{\nabla} )
    p_\eta(\boldsymbol{x},t)~,
\end{equation}
where the subscript $\eta$ denotes the non-white $\boldsymbol{\eta}_T$ noise.
Note that the probability current has an asymmetric diffusion matrix
$T \mathbf{G}^{-1}$, which is consistent with the singular correlation matrix in Eq.~\eqref{prop:co_mat} (explicitly shown in \cite{Chun2018emergence}) and is also
responsible for the curl flux mentioned above.

Nevertheless, the antisymmetric part of the diffusion matrix does not
contribute to the time evolution of  $p_\eta(\boldsymbol{x},t)$, as it is sandwiched between the
same gradient operators in the FP equation. Thus, the distribution $p_\eta(\boldsymbol{x},t)$ should be identical
to that for the (naive) overdamped system with the symmetric diffusion matrix
$(T_B/\gamma)\mathbf{I}$, which is the symmetric part of
the original diffusion matrix $T \mathbf{G}^{-1}$.

This naive overdamped Langevin equation can be simply obtained by setting $m=0$ in the
underdamped equation of Eq.~\eqref{eq:U_Langevin} as
\begin{equation}\label{eq:U_altLangevin}
    \dot{\boldsymbol{x}}(t) =
    \mathbf{G}^{-1} \boldsymbol{f} \qty(\boldsymbol{x} (t)) +
    \boldsymbol{\xi}_{T_B} (t)
\end{equation}
with a white Gaussian noise
$\boldsymbol{\xi}_{T_B}(t) = \mathbf{G}^{-1} \boldsymbol{\xi}_T(t)$
satisfying $\expval*{\boldsymbol{\xi}_{T_B} (t) \boldsymbol{\xi}_{T_B}^\rT (t')}
= \mathbf{Z}_\mathrm{s}(t-t')=(2 T_B/\gamma) \mathbf{I} \delta(t-t')$.
The corresponding naive probability current is
\begin{equation}\label{temp:o_pro_n_current}
    \boldsymbol{J}_\xi (\boldsymbol{x},t) = \qty(\mathbf{G}^{-1}
     \boldsymbol{f}(\boldsymbol{x}) -\frac{T_B}{\gamma} \boldsymbol{\nabla} )
    p_\xi(\boldsymbol{x},t)~,
\end{equation}
We will refer this naive dynamics as $\xi$-dynamics,
whereas the original overdamped dynamics with the non-white noise as $\eta$-dynamics.

The equivalence of the distribution functions for both dynamics as $p_\eta(\boldsymbol{x},t)=p_\xi(\boldsymbol{x},t)$
guarantees that
the average of any position-dependent observable (internal energy, system Shannon entropy, etc)
does not discriminate the true overdamped $\eta$-dynamics and the naive (incorrect) $\xi$-dynamics.
However, the average of path-dependent observables such as currents
could depend on the antisymmetric part of the diffusion matrix.
One specific example has been already reported in Ref.~\cite{Chun2018emergence} by an explicit calculation of the average work currents
for two different dynamics for an exactly solvable model, which turn out to be
clearly different from each other. Therefore, one should be careful in discussing the TUR involving the current
average as well as its variance in the presence of a magnetic Lorentz force.

\section{Extended FP equation and alternative dynamical observable}
\label{sec:III}

Consider a general accumulated current $\Phi$ in the overdamped regime as
\begin{equation}\label{def:neq_current}
    \Phi (\Gamma) = \int_0^t dt' \boldsymbol{\Lambda}^\rT (\boldsymbol{x}(t')) \circ
    \dot{\boldsymbol{x}} (t')~,
\end{equation}
where $\Gamma=\{\boldsymbol{x}(t')|t' \in (0,t)\}$ denotes
a trajectory in the state space, $\boldsymbol{\Lambda} (\boldsymbol{x})$
is an arbitrary state-dependent vector (`weight' function), and $\circ$ represents
the Stratonovich product.

It is convenient to consider the joint distribution
$\hat{p}(\boldsymbol{x},\Phi,t)$ for the position and the current,
from which the current average and its fluctuations
can be easily calculated.
As an additional stochastic variable, the current satisfies
\begin{align}\label{eq:current_rate}
\dot{\Phi}(t)=\boldsymbol{\Lambda}^\rT (\boldsymbol{x}(t))\circ
    \dot{\boldsymbol{x}} (t)~.
\end{align}
Together with Eq.~\eqref{eq:U_altLangevin},
one can derive the extended FP equation for the naive $\xi$-dynamics as
\begin{align}
\label{eq:exo_FP_eq_zeta1}
    \partial_t \hat{p}_\xi(\boldsymbol{x},\Phi,t)=\hat{\mathcal{L}}_{\xi,\Phi} \hat{p}_\xi(\boldsymbol{x},\Phi,t)~,
\end{align}
with the FP operator
\begin{align}\label{eq:exo_FP_eq_zeta2}
\hat{\mathcal{L}}_{\xi,\Phi}
    = - \tilde{\boldsymbol{\nabla}}_\Phi^\rT \cdot \left(\mathbf{G}^{-1} \boldsymbol{f} (\boldsymbol{x})
    - \frac{T_B}{\gamma} \tilde{\boldsymbol{\nabla}}_\Phi\right)
\end{align}
where $\hat{\cdot}$ denotes quantities and operators in the extended phase space and $\tilde{\boldsymbol{\nabla}}_\Phi = \boldsymbol{\nabla} + \partial_\Phi \boldsymbol{\Lambda}(\boldsymbol{x})$ is a {\em tilted} gradient
operator. For detailed derivation, see Appendix~\ref{sec:app_B}. By integrating out over $\Phi$ in Eq.~\eqref{eq:exo_FP_eq_zeta1},
one can recover the ordinary FP equation with the probability current given in Eq.~\eqref{temp:o_pro_n_current}.

Now, let us derive the extended FP equation for the original $\eta$-dynamics.
As the $\boldsymbol{\eta}_T$ noise is non-white, it is not straightforward to derive the extended FP operator directly.
Thus, we again go back to the underdamped version and
take the small-mass expansion. After a lengthy but straightforward calculation (shown in Appendix~\ref{sec:app_C}), we obtain
\begin{align}
\label{eq:exo_FP_eq_eta1}
    \partial_t \hat{p}_\eta(\boldsymbol{x},\Phi,t)=\hat{\mathcal{L}}_{\eta,\Phi} \hat{p}_\eta(\boldsymbol{x},\Phi,t)~,
\end{align}
with
\begin{align}\label{eq:exo_FP_eq_eta2}
\hat{\mathcal{L}}_{\eta,\Phi}
    = - \tilde{\boldsymbol{\nabla}}_\Phi^\rT \mathbf{G}^{-1} \left(\boldsymbol{f} (\boldsymbol{x})
    - T\tilde{\boldsymbol{\nabla}}_\Phi\right)
    =\hat{\mathcal{L}}_{\xi,\Phi} - \partial_\Phi \varphi(\boldsymbol{x})~,
\end{align}
where the scalar function
\begin{align}\label{eq:phi}
\varphi(\boldsymbol{x}) = - T \boldsymbol{\nabla}^\rT \mathbf{G}_\mathrm{a}^{-1} \boldsymbol{\Lambda} (\boldsymbol{x})
\end{align}
with $\mathbf{G}_\mathrm{a}^{-1} = (\mathbf{G}^{-1} - (\mathbf{G}^{-1})^\rT)/2$.
Note that the gradient operator in Eq.~\eqref{eq:phi} works only inside the scalar function (not on the distribution function).
This scalar function is originated from the antisymmetric part of the diffusion matrix and contributes to the time evolution of the joint distribution, in contrast to the ordinary distribution case. Thus, $\hat{p}_\eta(\boldsymbol{x},\Phi,t)\neq \hat{p}_\xi(\boldsymbol{x},\Phi,t)$ and thus $\langle {\cal A}(\Phi)\rangle_\eta \neq \langle {\cal A}(\Phi)\rangle_\xi$ for an arbitrary function ${\cal A}(\Phi)$.

Our key observation is that the extended FP operator $\hat{\mathcal{L}}_{\eta,\Phi}$ has the same mathematical structure with the time evolution operator $\hat{\mathcal{L}}_{\xi,\Psi}$ for the joint distribution $\hat{p}_\xi(\boldsymbol{x},\Psi,t)$ with the alternative dynamic observable
\begin{align}\label{def:Psi}
    \Psi (\Gamma) = \Phi (\Gamma)
    + \int_0^t d t' \varphi(\boldsymbol{x}(t'))
\end{align}
in the $\xi$-dynamics. In short, we find the operator correspondence relation
\begin{equation}\label{eq:op_correspond}
    \hat{\mathcal{L}}_{\eta,\Phi}  = \left . \hat{\mathcal{L}}_{\xi,\Psi}\right |_{\Psi\rightarrow\Phi}~,
\end{equation}
which is easily verified from Eq.~\eqref{eqA:PsiFP} in Appendix~\ref{sec:app_B}
for the explicit derivation of $\hat{\mathcal{L}}_{\xi,\Psi}$ and Eq.~\eqref{eq:exo_FP_eq_eta2}.
Note that the alternative dynamic observable $\Psi$ and the original current $\Phi$ differ by the accumulated state-dependent observable $\varphi$, meaning that $\Psi$ is not antisymmetric under time reversal in contrast to $\Phi$.

\begin{figure}
\includegraphics*[width=\columnwidth]{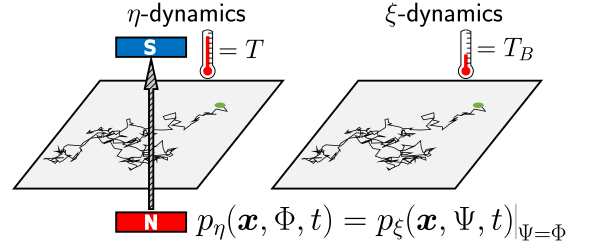}
\caption{
A schematic description representing the relation between
the original $\eta$-dynamics and the naive
$\xi$-dynamics.
}
\label{fig:mapping}
\end{figure}

The correspondence relation between the extended FP operators implies that
the random variables $\Phi$ in the $\eta$-dynamics and $\Psi$ in the $\xi$-dynamics
should be equivalent in their distributions at any time during  the evolution, starting with
the same initial distribution.
In other words, we can identify the equivalence relations as
\begin{align}\label{eq:correspond_rel}
\hat{p}_\eta(\boldsymbol{x},\Phi,t) = \left .\hat{p}_\xi(\boldsymbol{x},\Psi,t) \right |_{\Psi \rightarrow \Phi}~\textrm{and}~
\langle {\mathcal A}(\Phi)\rangle_\eta = \langle {\mathcal A}(\Psi)\rangle_\xi~.
\end{align}
Using these relations, we can avoid complex calculations for the averages,
caused by the singular nature of the non-white noise in the original $\eta$-dynamics.
Figure~\ref{fig:mapping} shows the schematic description
of this relation.
For example, the current average and its variance in the $\eta$-dynamics are identical to
those of the alternative dynamic observable in the $\xi$-dynamics as
\begin{align}\label{eq:Phi_Var}
\langle\Phi\rangle_\eta=\langle\Psi\rangle_\xi~~~ \textrm{and}~~
\textrm{Var}_\eta[\Phi]=\textrm{Var}_\xi[\Psi]~,
\end{align}
which are useful to calculate the TUR factor ${\cal Q}_\Phi$ for the original $\eta$-dynamics.

The time evolution of the alternative dynamic observable $\Psi$ reads from Eq.~\eqref{def:Psi}
\begin{align}\label{eq:Phi_Psi}
\dot{\Psi}(t)=\boldsymbol{\Lambda}^\rT (\boldsymbol{x}(t))\circ
\dot{\boldsymbol{x}} (t) +\varphi(\boldsymbol{x}(t))~,
\end{align}
yielding the relation between the average current rates
\begin{align}\label{eq:Phi_diff}
\expval*{\dot{\Phi}}_\eta=\expval*{\dot{\Psi}}_\xi
=\expval*{\dot{\Phi}}_\xi+\langle\varphi\rangle_\xi~.
\end{align}
It is worth mentioning that the non-zero average scalar term $\langle\varphi\rangle_\xi$ represents the difference of two different averages of the current $\Phi$.

%%  Fig. 2  %%%%%%%%%%%%%%%%%%%%%%%%%%%%%%%%%%%%%%%%%%
%%%%%%%%%%%%%%%%%%%%%%%%%%%%%%%%%%%%%%%%%%%%%%%%%%%%%%

As in the traditional stochastic thermodynamics like in the $\xi$-dynamics, the average current rate in the $\eta$-dynamics is also written
in an integral form involving the probability current (shown in Appendix~\ref{sec:app_D})
as
\begin{align}\label{eq:ave_Phi}
 \langle{\dot{\Phi}}\rangle_\eta
    = \int d \boldsymbol{x}~ \boldsymbol{\Lambda}^\rT(\boldsymbol{x}) \cdot
        \boldsymbol{J}_\eta (\boldsymbol{x},t)~.
\end{align}
The average rate of $\Psi$ in the $\xi$-dynamics is also written in a similar form, using
Eq.~\eqref{eq:Phi_Psi}, as
\begin{align}
 \langle{\dot{\Psi}}\rangle_\xi
    &= \int d \boldsymbol{x}~ \qty( \boldsymbol{\Lambda}^\rT(\boldsymbol{x}) \cdot
    \boldsymbol{J}_\xi (\boldsymbol{x},t)
    + \varphi(\boldsymbol{x}) p_\xi (\boldsymbol{x},t) )     \\
    &\equiv \int d \boldsymbol{x}~ \boldsymbol{\Lambda}^\rT(\boldsymbol{x}) \cdot
        \boldsymbol{\mathcal{J}}_\xi (\boldsymbol{x},t)\label{eq:Psi_eff_cur}
\end{align}
with the {\em effective} probability current
\begin{align}
    \boldsymbol{\mathcal{J}}_\xi (\boldsymbol{x},t)
    &= \boldsymbol{J}_\xi (\boldsymbol{x},t)-T \mathbf{G}^{-1}_\textrm{a} \boldsymbol{\nabla} p_\xi (\boldsymbol{x},t) \label{eq:eff_cur1}\\
    &= \mathbf{G}^{-1}  \qty( \boldsymbol{f} (\boldsymbol{x})
    - T \boldsymbol{\nabla} ) p_\xi (\boldsymbol{x},t)~,\label{eq:eff_cur2}
\end{align}
where we used Eq.~\eqref{eq:phi} and the relation $\int d \boldsymbol{x}~ (\mathbf{G}^{-1}_\textrm{a} \boldsymbol{\nabla})^\rT \cdot\boldsymbol{\Lambda}(\boldsymbol{x}) p_\xi (\boldsymbol{x},t)=0$.
Comparing this with $\boldsymbol{J}_\eta (\boldsymbol{x}, t)$ in Eq.~\eqref{temp:o_pro_current}, we find the relation of
\begin{align}\label{eq:JJ_rel}
\boldsymbol{\mathcal{J}}_\xi (\boldsymbol{x},t)=
 \boldsymbol{J}_\eta (\boldsymbol{x},t)~,
\end{align}
with the distribution equivalence of $p_\eta(\boldsymbol{x},t)=p_\xi(\boldsymbol{x},t)$.
This relation along with Eqs.~\eqref{eq:ave_Phi} and \eqref{eq:Psi_eff_cur} confirms Eq.~\eqref{eq:Phi_diff}.

To discuss the TUR, we consider the steady-state entropy production rate
for the $\eta$-dynamics. Its stochastic version in the steady state can be written as
\begin{align}\label{eq:EP0}
\dot{S}^\textrm{tot}(t)=\boldsymbol{\Lambda}^\rT_S (\boldsymbol{x}(t))\circ
\dot{\boldsymbol{x}} (t)
~\textrm{with}~\boldsymbol{\Lambda}_S (\boldsymbol{x}) = \frac{\mathbf{G} \boldsymbol{J}_\eta^\mathrm{ss} (\boldsymbol{x})}{Tp_\eta^\mathrm{ss} (\boldsymbol{x})}
\end{align}
where the superscript `ss' denotes a steady-state quantity such as $\boldsymbol{J}_\eta^\mathrm{ss} (\boldsymbol{x}) = \mathbf{G}^{-1} (\boldsymbol{f}(\boldsymbol{x}) -T \boldsymbol{\nabla} ) p_\eta^\mathrm{ss}(\boldsymbol{x})$ with the steady-state distribution  $p_\eta^\mathrm{ss}(\boldsymbol{x})$ (see Appendix~\ref{sec:app_D}.3 for its derivation).
Using Eqs.~\eqref{eq:EP0} and \eqref{eq:ave_Phi},
we obtain the average steady-state rate
\begin{align}\label{eq:EP_total_zeta_1}
\langle \dot{S}^\textrm{tot}\rangle_\eta^\mathrm{ss}
&= \frac{\gamma}{T}
\int d\boldsymbol{x}~ \frac{|\boldsymbol{J}_\eta^\mathrm{ss}(\boldsymbol{x})|^2}{p_\eta^\mathrm{ss}(\boldsymbol{x})}\\
&= \frac{\gamma}{T}
\int d\boldsymbol{x}~  \frac{| \boldsymbol{\mathcal{J}}_\xi^\mathrm{ss} (\boldsymbol{x}) |^2 }{p_\xi^\mathrm{ss} (\boldsymbol{x})}
\equiv \langle \dot{\Sigma}\rangle_\xi^\mathrm{ss}\label{eq:EP_total_zeta_2}
\end{align}
where the relation of Eq.~\eqref{eq:JJ_rel} is used
with $\boldsymbol{\mathcal{J}}_\xi^\mathrm{ss} (\boldsymbol{x}) = \mathbf{G}^{-1}  \qty( \boldsymbol{f} (\boldsymbol{x}) - T \boldsymbol{\nabla} ) p_\xi^\mathrm{ss} (\boldsymbol{x})$
and Eq.~\eqref{eq:Phi_diff} guarantees $\langle \dot{S}^\textrm{tot}\rangle_\eta^\mathrm{ss} =\langle \dot{{\Sigma}}\rangle_\xi^\mathrm{ss}$ with
the corresponding alternative dynamic variable
${\Sigma}$, defined as $\dot{\Sigma}(t)= \dot{S}^\textrm{tot}(t) +\varphi_S(\boldsymbol{x}(t))$ with
the scalar function $\varphi_S(\boldsymbol{x})=-T \boldsymbol{\nabla}^\rT \mathbf{G}_\mathrm{a}^{-1} \boldsymbol{\Lambda}_S(\boldsymbol{x})$ (see the explicit calculation  in Appendix~\ref{sec:app_D}.4).

Using Eqs.~\eqref{eq:Phi_Var} and \eqref{eq:EP_total_zeta_2}, the TUR factor ${\cal Q}_\Phi$
for the $\eta$-dynamics is now expressed  in terms of quantities in the $\xi$-dynamics as
\begin{align}\label{eq:TUR_xi}
{\cal Q}_\Phi &= \frac{\textrm{Var}_\eta^\mathrm{ss}[\Phi]}{{(\langle\Phi\rangle}_\eta^\mathrm{ss})^2} \langle\Delta S^\textrm{tot}\rangle^\mathrm{ss}_\eta
=\frac{\textrm{Var}_\xi^\mathrm{ss}[\Psi]}{{(\langle\Psi\rangle}_\xi^\mathrm{ss})^2} \langle\Delta \Sigma \rangle_\xi^\mathrm{ss}
\equiv {\cal R}_\Psi
\end{align}
with $\langle\Delta \Sigma \rangle_\xi^\mathrm{ss} = \int_0^t dt' \langle \dot{\Sigma} (t') \rangle_\xi^\mathrm{ss}$.
We refer to ${\cal R}_\Psi$ as the {\em alternative} TUR factor for $\Psi$ in the perspective of the $\xi$-dynamics.
In the normal overdamped $\xi$-dynamics, the conventional TUR should be valid for general irreversible currents
of the generic type in Eq.~\eqref{def:neq_current} with an arbitrary duration time $t$~\cite{Hasegawa2019uncertainty, Dechant2019multidimensional}.
However, our alternative dynamic observable $\Psi$ is not of the generic type, but is given by
the combination of the current $\Phi$ and the accumulated state-dependent variable $\varphi$.
Furthermore, $\Delta\Sigma$  is not the entropy production for the $\xi$-dynamics (see Eq.~\eqref{eqA:EP_Diff}
for the true entropy production). Thus, the alternative TUR factor $R_\Psi$ may violate the conventional TUR bound of $2k_\textrm{B}$.
In the next section, we show that this violation indeed occurs and derive a new
bound analytically.

\section{Modified TUR}\label{sec:IV}

The derivation of the conventional TUR is based on the observation that
the TUR is a special case of the Cramer-Rao (CR) inequality~\cite{Rao1945information,Cramer1999mathematical}.
Here, we take a similar but slightly different approach from the conventional one adopted in Refs.~\cite{Hasegawa2019uncertainty, Dechant2019multidimensional},
to derive a new lower bound for the alternative TUR factor ${\cal R}_\Psi$ in the $\xi$-dynamics.

The CR inequality generally provides  a lower bound of
the variance of any dynamic observable $O(\Gamma)$ in a stochastic process
with a parameter $\theta$ as
\begin{equation}\label{eq:CramerRao}
    \frac{\mathrm{Var}_\theta \left[ O  \right]}
    {|\partial_\theta \expval{O }_\theta |^2}
    I(\theta) \geq 1~,
\end{equation}
where
$\langle O\rangle_\theta = \int d\Gamma O(\Gamma) \mathcal{P}_\theta (\Gamma)$ with
the trajectory probability $\mathcal{P}_\theta (\Gamma)$ and
the Fisher information is defined as
\begin{align}\label{eq:Fisher_info}
I(\theta) = \expval{-\partial_\theta^2 \ln{\mathcal{P}_\theta}(\Gamma)}_\theta~.
\end{align}
If the dynamics of our interest is defined at $\theta=0$, it is crucial to
find an appropriate modified dynamics with a nonzero $\theta$, which yields
$\lim_{\theta\rightarrow 0} \partial_\theta  \expval{O}_\theta= \expval{O}_0$.
Then, the (alternative) TUR factor in Eq.~\eqref{eq:TUR_xi} naturally comes into the CR
inequality  at $\theta=0$.

We take the $\xi$-dynamics as the unperturbed one at $\theta=0$.
Similar to the perturbation term in the previous studies~\cite{Hasegawa2019uncertainty, Dechant2019multidimensional},
we consider  a linear perturbation on the force as
\begin{equation}\label{def:pert_force}
    \boldsymbol{f}_{\theta} (\boldsymbol{x}) = \boldsymbol{f} (\boldsymbol{x})
    + \theta \mathbf{G}
    \frac{\boldsymbol{\mathcal{J}}_\xi^\mathrm{ss} (\boldsymbol{x})}
    {p_\xi^\mathrm{ss}(\boldsymbol{x})}~.
\end{equation}
The steady-state distribution $p_\xi^\mathrm{ss}(\boldsymbol{x})$ satisfies $\boldsymbol{\nabla}^\rT \cdot \boldsymbol{J}_\xi^\mathrm{ss} (\boldsymbol{x})=0$
and we also find $\boldsymbol{\nabla}^\rT \cdot \boldsymbol{\mathcal{J}}_\xi^\mathrm{ss} (\boldsymbol{x})= 0$
as well from Eq.~\eqref{eq:eff_cur1}.
In the previous studies for the conventional TUR~\cite{Hasegawa2019uncertainty, Dechant2019multidimensional},
$\gamma \boldsymbol{J}_\xi^\mathrm{ss} (\boldsymbol{x})$ was used instead of $\mathbf{G} \boldsymbol{\mathcal{J}}_\xi^\mathrm{ss} (\boldsymbol{x})$
in the perturbation term of Eq.~\eqref{def:pert_force}.

It is easy to see that the perturbed steady-state distribution is identical to the unperturbed one,
i.e.~$p^\mathrm{ss}_{\xi,\theta}(\boldsymbol{x}) = p_\xi^\mathrm{ss}(\boldsymbol{x})$, as
the perturbed steady-state probability current
\begin{equation}
    \boldsymbol{J}^\mathrm{ss}_{\xi,\theta}(\boldsymbol{x})
    = \qty( \mathbf{G}^{-1}  \boldsymbol{f}_\theta (\boldsymbol{x})
    - \frac{T_B}{\gamma} \boldsymbol{\nabla} )
    p^\mathrm{ss}_{\xi,\theta}(\boldsymbol{x})
\end{equation}
satisfies the steady-state condition of $\boldsymbol{\nabla}^\rT \cdot \boldsymbol{J}_{\xi,\theta}^\mathrm{ss} (\boldsymbol{x})=0$.
The perturbed effective probability current in the steady state becomes
\begin{equation}
    \boldsymbol{\mathcal{J}}_{\xi,\theta}^\mathrm{ss} (\boldsymbol{x})
    = \mathbf{G}^{-1}  \qty( \boldsymbol{f}_\theta (\boldsymbol{x})
    - T \boldsymbol{\nabla} ) p_{\xi,\theta}^\mathrm{ss} (\boldsymbol{x})
    = ( 1 + \theta ) \boldsymbol{\mathcal{J}}_\xi^\mathrm{ss}
    (\boldsymbol{x}),
\end{equation}
which yields
$\expval*{\dot{\Psi}}_{\xi,\theta}^\mathrm{ss} =
(1 + \theta) \expval*{\dot{\Psi}}_\xi^\mathrm{ss}$ from Eq.~\eqref{eq:Psi_eff_cur},
thus we find
\begin{align}\label{eq:partial_theta}
\partial_\theta \expval*{\dot{\Psi}}_{\xi,\theta}^\mathrm{ss}=\expval*{\dot{\Psi}}_\xi^\mathrm{ss}~.
\end{align}

The remaining task is to express the Fisher information $I(\theta)$ at $\theta=0$
in terms of $\expval*{\Delta \Sigma}_\xi^\mathrm{ss}$.
Using the Onsager-Machlup
theory~\cite{Onsager1953fluctuations} in Eq.~\eqref{eq:U_altLangevin}, the trajectory probability is written as
\begin{equation}\label{eq:path_prob}
    \mathcal{P}_\theta [\Gamma]
    = p_{\xi,\theta}^\mathrm{ss} (\boldsymbol{x} (0))
    \mathcal{N}_\theta
    e^{ - \frac{\gamma}{4 T_B}
    \int_0^t dt'
    \qty[ \dot{\boldsymbol{x}} (t') - \mathbf{G}^{-1}
    \boldsymbol{f}_\theta (\boldsymbol{x}(t')) ]^2}
\end{equation}
where the initial probability $p_{\xi,\theta}^\mathrm{ss} (\boldsymbol{x} (0))$ and the normalization factor $\mathcal{N}_\theta$
are $\theta$-independent. Then, one can easily find the explicit form of the Fisher information from Eq.~\eqref{eq:Fisher_info} as
\begin{equation}\label{eq:o_F_info}
    I(\theta) = \frac{t\gamma}{2T_B}
    \int d\boldsymbol{x}
    \frac{
    | \boldsymbol{\mathcal{J}}_\xi^\mathrm{ss} (\boldsymbol{x}) |^2
    }{p_\xi^\mathrm{ss} (\boldsymbol{x})}~,
\end{equation}
which is $\theta$-independent. Using Eq.~\eqref{eq:EP_total_zeta_2},
we find
\begin{align}\label{eq:Itheta}
I(\theta) = t \frac{T}{2T_B} \langle \dot{\Sigma}\rangle_\xi^\mathrm{ss}=
\frac{T}{2T_B} \langle\Delta \Sigma\rangle_\xi^\mathrm{ss}~.
\end{align}

Using Eqs.~\eqref{eq:TUR_xi}, \eqref{eq:CramerRao}, \eqref{eq:partial_theta} and \eqref{eq:Itheta} along with Eq.~\eqref{eq:TB} and setting $\theta=0$, we finally obtain the modified TUR for the overdamped $\eta$-dynamics as
\begin{align}\label{eq:main_result}
{\cal Q}_\Phi=
\frac{\mathrm{Var}_\eta^\mathrm{ss} \left[ \Phi  \right]}
    {(\expval{\Phi }_\eta^\mathrm{ss})^2}
    \langle\Delta S^\textrm{tot}\rangle_\eta^\mathrm{ss}={\cal R}_\Psi
    \ge 2k_\textrm{B} \frac{T_B}{T}
    =\frac{2k_\textrm{B}}{1+(B/\gamma)^2},
\end{align}
which is the main result of our paper. Note that we restored the Boltzmann constant $k_\textrm{B}$ here.
The result shows that the Lorentz magnetic force always lowers the threshold of the TUR factor,
which weakens the trade-off constraint. Obviously, the conventional TUR is recovered at $B=0$.
Interestingly, Chun {\it et al}~\cite{Chun2019effect} found the same form of the TUR lower bound in an exactly solvable linear model
for the work and heat current in the long-time ($t\rightarrow\infty$) limit. Our result applies to
a general current in a general nonlinear system for an arbitrary duration time. In the next section,
we derive the TUR factors exactly for the linear model for a finite duration time and confirm the validity of our
modified TUR.

The equality condition of the modified TUR can be
determined by the equality condition of the Cramer-Rao inequality
in Eq.~\eqref{eq:CramerRao} as
\begin{align}
\partial_\theta\ln  {\cal P}_\theta (\Gamma) \propto O(\Gamma)-\langle O\rangle_\theta~
~~\textrm{for any}~ \Gamma~.
\end{align}
After a straightforward calculation following a similar procedure in Ref.~\cite{Hasegawa2019uncertainty}, we find
that the equality holds with the two constraints as
\begin{equation}\label{eq:eq_condition1}
    \boldsymbol{\Lambda}(\boldsymbol{x}) =\mu
    \frac{\boldsymbol{J}_\eta^\mathrm{ss}(\boldsymbol{x})}{p_\eta^\mathrm{ss}(\boldsymbol{x})}
\end{equation}
with an arbitrary constant $\mu$, and
\begin{equation}\label{eq:eq_condition2}
\langle \dot{S}^\textrm{tot}\rangle_\eta^\mathrm{ss}
    = \frac{\gamma}{T} \qty |
    \frac{\boldsymbol{J}_\eta^\mathrm{ss}(\boldsymbol{x})}
    {p_\eta^\mathrm{ss}(\boldsymbol{x})}
    |^2
    - \gamma \frac{\boldsymbol{\nabla}^\rT
    \mathbf{G}_a^{-1} \boldsymbol{J}_\eta^\mathrm{ss} (\boldsymbol{x})}
    {p_\eta^\mathrm{ss}(\boldsymbol{x})}~.
\end{equation}
Note that the entropy production current automatically satisfies the first constraint
for $B=0$, but not for $B\neq 0$ (see Eq.~\eqref{eq:EP0}). Other currents
may satisfy the first constraint for some specific dynamics
(see the example in the next section), but not in general.
The second constraint is independent of the choice of currents, $\boldsymbol{\Lambda}(\boldsymbol{x})$,
and more interesting.
For $B = 0$, the second term in the right hand side of Eq.~\eqref{eq:eq_condition2} vanishes,
thus it reduces to the equality condition for the conventional TUR~\cite{Hasegawa2019uncertainty},
which requires {\em uniform} local entropy production.
This second constraint becomes trivially satisfied
in the equilibrium limit where $\boldsymbol{J}_\eta^\mathrm{ss} (\boldsymbol{x})$ vanishes (detailed balance).
For $B \neq 0$, however, the second term exists and in fact dominates near equilibrium,
as it is of the first order in $\boldsymbol{J}_\eta^\mathrm{ss} (\boldsymbol{x})$, while the others, including $\langle \dot{S}^\textrm{tot}\rangle_\eta^\mathrm{ss}$, are of the second order.
Hence, the equality condition is violated even in the equilibrium limit. Instead,
the second constraint can be satisfied by tuning $B$ (thus, $\mathbf{G}_a^{-1} $)
to make the second term also be of the second order in $\boldsymbol{J}_\eta^\mathrm{ss} (\boldsymbol{x})$.
In the next section, we will show this example explicitly, where
the equality of the modified TUR holds out of equilibrium.

It is straightforward to extend our modified TUR to the $\eta$-dynamics with an arbitrary initial state.
This generalization in the normal overdamped $\xi$-dynamics was reported recently in~\cite{Dechant2018current, Liu2020thermodynamic}.
We also study this generalization in the $\eta$-dynamics and find
the modified TUR  with an arbitrary initial state for a finite
duration time $t$ as
\begin{align}\label{eq:TUR_arb}
\frac{\mathrm{Var}_\eta \left[ \Phi  \right]}
    {\expval*{t\dot{\Phi}(t)}_\eta^2}
    \langle\Delta S^\textrm{tot}\rangle_\eta
    \ge 2k_\textrm{B} \frac{T_B}{T}~,
\end{align}
where $\dot{\Phi}(t)$ is the current at the final time $t$ of duration and
Eq.~\eqref{eq:main_result}  is recovered in the steady state.
The derivation details are given in Appendix~\ref{sec:app_E}.

A generalization to higher dimensional cases is also possible. The overdamped motion in the perpendicular plane
to the magnetic field follows the same behavior discussed before with the effective temperature $T_B$. In the parallel plane, the particle does not feel the Lorentz force and the normal overdamped motion is expected with the temperature $T$.
The contribution to the Fisher information in Eq.~\eqref{eq:o_F_info} from this normal overdamped motion in the parallel plane is
smaller as $T>T_B$, thus the total Fisher information is smaller than the entropy production of the $\eta$-dynamics in
Eq.~\eqref{eq:EP_total_zeta_1}. Therefore, our main result still holds in higher dimensions, even though the TUR equality condition can not be satisfied
for non-zero $B$. For details, see Appendix~\ref{sec:app_F}.

\section{Example}\label{sec:V}

%%  Fig. 3  %%%%%%%%%%%%%%%%%%%%%%%%%%%%%%%%%%%%%%%%%%
\begin{figure*}
\includegraphics*[width=2\columnwidth]{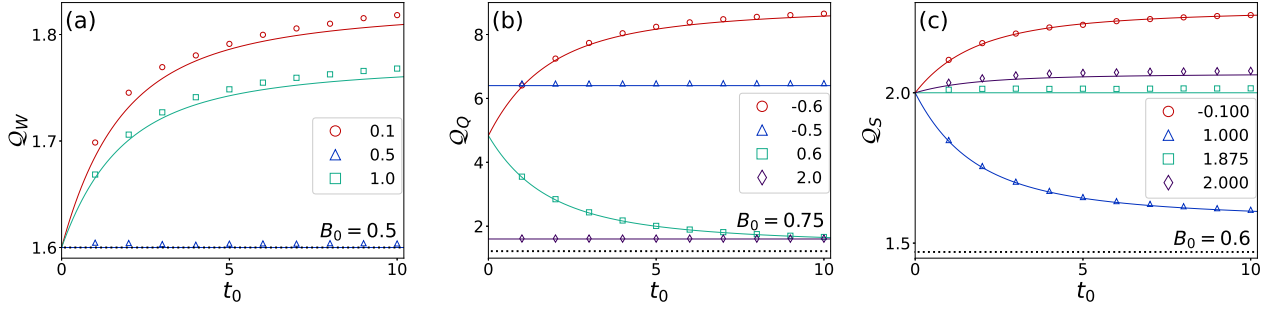}
\caption{
Duration-time-dependent behavior of the TUR factors (a) for work $\mathcal{Q}_W$, (b) for heat $\mathcal{Q}_Q$, and (c) for the entropy production $\mathcal{Q}_S$. The symbols represent the numerical data from the underdamped dynamics with $m_0 = 0.001$. Lines represent the analytic results obtained in the overdamped $\eta$-dynamics. Black dotted lines indicate the modified TUR bound in Eq.~\eqref{eq:main_result}. We fixed (a) $B_0 = 0.5$ with several values of $\epsilon_0 = 0.1$ (red circle), $0.5$ (blue triangle), and $1.0$. (cyan square) for work, (b) $B_0 = 0.75$ with $\epsilon_0 = -0.6$ (red circle), $-0.5$ (blue triangle), $0.6$ (cyan square), and $2.0$ (violet diamond) for heat, and (c) $B_0 = 0.6$ and $\epsilon_0 = -0.1$ (red circle), $1.0$ (blue triangle), $1.875$ (cyan square), and $2.0$ (violet diamond) for entropy production.
}
\label{fig:Qfactor}
\end{figure*}
%%%%%%%%%%%%%%%%%%%%%%%%%%%%%%%%%%%%%%%%%%%%%%%%%%%%%%

To confirm our main result,
we consider an analytically solvable system
with a linear force as
\begin{align}\label{eq:Fx}
\boldsymbol{f} (\boldsymbol{x}) = - \mathbf{F}\boldsymbol{x}~\quad\textrm{with}~\quad    \mathbf{F} =
    \begin{pmatrix}
        k & - \epsilon\\
        \epsilon & k
    \end{pmatrix}~.
\end{align}
The diagonal elements represent
a force applied by a harmonic potential with
spring constant $k$, while the off-diagonal elements indicate
a nonequilibrium driving force with strength $\epsilon$.
This system has been investigated in various
studies~\cite{Lee2019thermodynamic,
Chun2019effect, Chun2018emergence, Lee2019nonequilibrium, Noh2012fluctuation, Park2016efficiency},
because it is analytically solvable.
In the remaining text, we will omit the subscript $\eta$
for brevity.

We consider two accumulated currents of work $W$ done on the system and heat $Q$ dissipated  into the reservoir, which are
defined in Eq.~\eqref{def:neq_current} with weight finctions
\begin{align}\label{eq:ex_current_w}
    \boldsymbol{\Lambda}_W (\boldsymbol{x}) &=\mathbf{W} \boldsymbol{x} \quad\textrm{with}~
    \mathbf{W} =
        \begin{pmatrix}
        0 & \epsilon\\
        - \epsilon & 0
   \end{pmatrix} \quad\textrm{for work}~W~, \\
    \boldsymbol{\Lambda}_Q (\boldsymbol{x}) &=\mathbf{Q} \boldsymbol{x} \quad\textrm{with}~\mathbf{Q}=-\mathbf{F}
   \quad\textrm{for heat}~Q~.\label{eq:ex_current_q}
\end{align}

Since the model basically belongs to the Ornstein-Uhlenbeck process~\cite{Gardiner:2010stochastic,Risken:1996FokkerPlanck,Lee2020exactly},
the steady-state distribution $p^\mathrm{ss}(\boldsymbol{x})$
should be Gaussian as
\begin{equation}\label{eq:ss_Gaussian}
    p^\mathrm{ss} (\boldsymbol{x})
    = \frac{1}{\sqrt{\det (2 \pi \mathbf{C})}}
    \exp{ - \frac{1}{2} \boldsymbol{x}^T \mathbf{C}^{-1}  \boldsymbol{x}}
\end{equation}
with the covariance matrix $\mathbf{C}=\langle \boldsymbol{x} \boldsymbol{x}^\rT\rangle$.
From the steady-state condition, we can easily find in Eq.~\eqref{eqA:Cov} as
\begin{equation}
    \mathbf{C} = \frac{\gamma T}{\gamma k + \epsilon B}
    \mathbf{I}~,
\end{equation}
which is stable for $\gamma k + \epsilon B > 0$.
The steady-state probability current
is then obtained from Eq.~\eqref{temp:o_pro_current} as
\begin{equation}\label{eq:ex_ss_prob_current}
    \boldsymbol{J}^\mathrm{ss} (\boldsymbol{x})
    = \frac{1}{\gamma}
    \mathbf{W} \boldsymbol{x} p^\mathrm{ss} (\boldsymbol{x})~.
\end{equation}

The steady-state averages of work $W$ and heat $Q$
are calculated by Eq.~\eqref{eq:ave_Phi}  as
\begin{align}
\expval*{W}^\textrm{ss} &=t\int d\boldsymbol{x}~ \boldsymbol{\Lambda}_W^\rT (\boldsymbol{x}) \cdot \boldsymbol{J}^\textrm{ss}(\boldsymbol{x}) \\
&=t\frac{2 \epsilon^2 T}{\gamma k + \epsilon B}  = \expval*{Q}^\textrm{ss}~,
\end{align}
which agrees with the result in~\cite{Chun2018emergence}.
The scalar function in Eq.~\eqref{eq:phi} becomes a constant as $\varphi(\boldsymbol{x})=2\epsilon BT/(\gamma^2+B^2)$
for both currents, which represents the difference between the true and the naive average
 in Eq.~\eqref{eq:Phi_diff}.

One can also consider the total entropy production $\Delta S^\textrm{tot}$ with
$\boldsymbol{\Lambda}_S (\boldsymbol{x})= \mathbf{G}\boldsymbol{J}^\textrm{ss}(\boldsymbol{x}) /[Tp^\textrm{ss}(\boldsymbol{x})]$ in the steady state
(see Eq.~\eqref{eq:EP0}). Using Eq.~\eqref{eq:ex_ss_prob_current}, we find the weight function as
\begin{align}
\boldsymbol{\Lambda}_S (\boldsymbol{x})= \mathbf{S} \boldsymbol{x} \quad\textrm{with}~\mathbf{S}={\mathbf{G}\mathbf{W}}/(T\gamma)
\quad\textrm{for }~\Delta S^\textrm{tot}~,
\end{align}
from which we obtain $\langle\Delta S^\textrm{tot}\rangle^\textrm{ss}= \expval*{Q}^\textrm{ss}/T$ and the corresponding scalar function
$\varphi_S(\boldsymbol{x})=2\epsilon B/(\gamma^2+B^2)$,
as expected.

The most complex task is to find the variance
of work, heat, and enropy production. From the extended FP equation
of $\hat{p}(\boldsymbol{x},\Phi,t)$ in Eqs.~\eqref{eq:exo_FP_eq_eta1} and \eqref{eq:exo_FP_eq_eta2},
we derive the formal solution
for the variance in terms of auxiliary matrices (see  Appendix~\ref{sec:app_G}).
Finding the explicit forms of the auxiliary matrices,
we obtain the variance and the TUR factor in the steady state
with a finite duration time $t$.

First, the TUR factor for work is given as
\begin{equation}\label{eq:Q_work}
    \mathcal{Q}_W
    = \frac{2 (1 + \epsilon_0^2)}
    {(1 + \epsilon_0 B_0)^2} -
    \frac{ 2(B_0 - \epsilon_0)^2}
    {(1 + B_0^2)(1 + \epsilon_0 B_0)^2}
    \frac{1 - e^{-t_0}}{t_0}~,
\end{equation}
where the dimensionless parameters are
\begin{align}
\epsilon_0 = \frac{\epsilon}{k}, ~ B_0 = \frac{B}{\gamma}, ~
t_0 = \frac{2t}{\tau}\qty(\frac{1+\epsilon_0 B_0}{1+B_0^2})
\end{align}
with $\tau=\gamma/k$ and the stability condition becomes $1+ \epsilon_0 B_0>0$.
It monotonically increases with $t_0$ with the minimum,
$\mathcal{Q}_W^\textrm{m}=2/(1 + B_0^2)$ at $t_0=0^+$, which
is exactly the same as the TUR lower bound in Eq.~\eqref{eq:main_result}.
This confirms the validity of our modified TUR.

Eq.~\eqref{eq:ex_ss_prob_current} indicates that the
work current obeys the first equality constraint in Eq.~\eqref{eq:eq_condition1}.
In addition, it is easy to show that the second constraint in  Eq.~\eqref{eq:eq_condition2}
is satisfied at $\epsilon_0 = B_0$, where the TUR equality should hold.
From Eq.~\eqref{eq:Q_work} with this condition, the TUR factor takes its minimum $\mathcal{Q}_W=\mathcal{Q}_W^\textrm{m}$,
which is independent of $t_0$. Thus, the TUR lower bound is achieved out of equilibrium with $\epsilon_0\neq0$.
This may be useful in inferring the entropy production by measuring current statistics~\cite{li2019quantifying,manikandan2020inferring,van2020entropy,otsubo2020estimating}, which should be
more feasible in nonequilibrium.

We check our analytic results numerically by simulating the underdamped dynamics
with a very small mass $m$. Using a generalized velocity-Verlet algorithm~\cite{Vanden2006second},
we perform numerical integrations of
the underdamped Langevin equation in
Eq.~\eqref{eq:U_Langevin} with dimensionless mass
$m_0 = k m /\gamma^2 = 0.001$.
Numerical data are plotted in Fig.~\ref{fig:Qfactor} (a)
for several values of $\epsilon_0$
with $B_0 = 0.5$ and $k = \gamma = T = 1$.
We used $10^7$ samples to evaluate
the average and the variance of work as well as the average entropy production.
The numerical data are in an excellent agreement with
Eq.~\eqref{eq:Q_work} with a small deviation
due to finite-mass corrections. Note that ${\cal Q}_W$ takes the modified TUR bound value, independent of $t_0$
at $\epsilon_0=B_0=0.5$, as expected.

We also find the TUR factor for heat as
\begin{equation}\label{eq:Q_heat}
    \mathcal{Q}_Q
    = \frac{2 (1 + \epsilon_0^2)}
    {(1 + \epsilon_0 B_0)^2}
    - \frac{2 (1 + \epsilon_0^2)
    (\epsilon^2_0 - 1 - 2 \epsilon_0 B_0)}
    {\epsilon^2_0 (1 + B_0^2)(1 + \epsilon_0 B_0)^2}
    \frac{1 - e^{-t_0}}{t_0}~,
\end{equation}
which monotonically increases with $t_0$ for $\epsilon_0^2 - 1 - 2 \epsilon_0 B_0>0$
and decreases for $\epsilon_0^2 - 1 - 2 \epsilon_0 B_0<0$.
Interestingly, the heat TUR factor $\mathcal{Q}_Q $
is always larger than the work TUR factor $\mathcal{Q}_W$ as
\begin{equation}\label{eq:diff_QQ_QW}
    \mathcal{Q}_Q - \mathcal{Q}_W =
    \frac{1}{\epsilon_0^2(1+B_0^2)}
    \frac{1 - e^{-t_0}}{t_0} \ge 0~,
\end{equation}
with the equality in the $t_0\rightarrow\infty$ limit.
Thus, $\mathcal{Q}_Q$ also
satisfies the modified TUR. The equivalence of $\mathcal{Q}_W=\mathcal{Q}_Q$
in the $t_0\rightarrow\infty$ limit was also reported in~\cite{Chun2019effect}.
Numerical data for $\mathcal{Q}_Q$ also agree with the analytic results
very well, shown in Fig.~\ref{fig:Qfactor} (b).
The TUR equality can not be achieved for heat current, as the
first equality constraint in Eq.~\eqref{eq:eq_condition1} is not satisfied.

The TUR factor for the entropy production (also called the entropy Fano factor~\cite{Pigolotti2017generic}) is
\begin{equation}\label{eq:Q_EP}
    \mathcal{Q}_S
    = \frac{2 (1 + \epsilon_0^2)}
    {(1 + \epsilon_0 B_0)^2}
    - \frac{2 \epsilon_0
    (\epsilon_0 - 2 B_0 - \epsilon_0 B_0^2)}
    {(1 + \epsilon_0 B_0)^2}
    \frac{1 - e^{-t_0}}{t_0}~,
\end{equation}
which is always larger than the work TUR factor as
\begin{equation}\label{eq:diff_QS_QW}
    \mathcal{Q}_S - \mathcal{Q}_W =
    \frac{2B_0^2}{1+B_0^2}
    \frac{1 - e^{-t_0}}{t_0} \ge 0~.
\end{equation}
Thus, $\mathcal{Q}_S$ also satisfies the modified TUR.
Interestingly, all three TUR factors
take the same value of $2(1+\epsilon_0^2)/(1 + \epsilon_0 B_0)^2$ in the limit of $t_0\rightarrow\infty$. The first equality constraint for $\mathcal{Q}_S$
is satisfied only for $B_0=0$, where the friction tensor $\mathbf{G}$ becomes proportional to
$\mathbf{I}$. As the second equality contraint of Eq.~\eqref{eq:eq_condition2}
is satisfied at $\epsilon_0=B_0$, we find the TUR equality only at $B_0=\epsilon_0=0$
with $\mathcal{Q}_S =2$.

It is also interesting to see that
$\lim_{t_0\rightarrow 0^+} \mathcal{Q}_S=2$, independent of the values of $B_0$ and $\epsilon_0$.
This may hint a possibility of the short-time TUR for inferring the entropy production~\cite{manikandan2020inferring}
even with the presence of a magnetic field. However, this does not come from the TUR equality constraint
and may not hold with a nonlinear force.
Figure~\ref{fig:Qfactor} (c) shows
this property of $\mathcal{Q}_S$ and the consistency between the numerical data and the analytic results.

Now, we make some remarks on near equilibrium  with small $\epsilon_0$,
where the TUR factors for work, heat, and the entropy production become
\begin{align}
    \mathcal{Q}_W
    &\approx 2 -
    \frac{2 B_0^2}{1 + B^2_0}
    \frac{1 - e^{-t_0}}{t_0}
    < 2 \\
    \mathcal{Q}_Q
    &\approx 2+ \frac{2 }{\epsilon_0^2(1 + B^2_0)}
     \frac{1 - e^{-t_0}}{t_0} >2\\
    \mathcal{Q}_S
    &\approx 2- 4\epsilon_0B_0 \qty(1-\frac{1 - e^{-t_0}}{t_0}) \approx 2~.
\end{align}
Even in the equilibrium limit, the work current always violates
the conventional TUR for finite $t_0$, which implies that any simulation
result should show a violation even with very small $\epsilon_0$
and very long $t_0$. In contrast, the heat current always satisfies
the conventional TUR for small $\epsilon_0$, but its finite-$\epsilon_0$ correction
is huge, implying that it is almost impossible to reach the
conventional TUR lower bound in any practical simulation near equilibrium.
In the case of the entropy production, the TUR factor approaches the conventional bound
from below for $B_0\epsilon_0>0$ and from above for  $B_0\epsilon_0<0$.
In the limit of $t_0\rightarrow\infty$ and
$\epsilon_0\rightarrow 0$, the TUR factor for all currents approaches 2 (the conventional TUR bound), independent of $B_0$.

Finally, we point out that a variant of the TUR
including the dynamic activity for the underdamped dynamics~\cite{Vu2019uncertainty, Lee2019thermodynamic}
is not useful in the overdamped limit, as the dynamic activity diverges with $m \rightarrow 0$
(see Eq.~(40) in Ref.~\cite{Lee2019thermodynamic} for this specific example).
We also note that the result from the linear response theory is not informative~\cite{Macieszczak2018unified}
as the Onsager matrix is symmetric in this example~\cite{Lee2020exactly}.

\section{conclusion}\label{sec:VI}

We present the modified TUR for the Langevin systems subject to a static magnetic field in the overdamped
limit. The system is described by the overdamped Langevin equation with a non-white singularly correlated noise.
By using the extended Fokker-Planck equation with a current variable in addition,
we find the alternative dynamics with the conventional white noise and an alternative dynamic observable, which generates
the extended distribution function equivalent to the original one.

Utilizing the Cramer-Rao inequality, we derive the modified TUR, which turns out to be surprisingly simple
with one scale factor on the conventional TUR bound.
This factor is given by the ratio of the reduced effective temperature
of the overdamped motion to the reservoir temperature. This TUR is universal
in the sense that the lower bound is independent of system parameters. We emphasize that
this TUR is valid for a finite duration time and for general currents.
Our modified TUR shows that the magnetic field lowers down the bound and the standard TUR
is recovered without the magnetic field. From the exactly solvable models, we confirm the validity of
our results. We also find that the TUR lower bound can be reached  out of equilibrium with the
magnetic field, which may be useful in inferring the entropy production by measuring non-entropic current statistics
in nonequilibrium. We also generalize our TUR for an arbitrary initial state.

Our analysis is limited to systems with isotropic thermal reservoirs.
It could be interesting to extend it to systems with anisotropic baths
such as microscopic heat engines with broken time-reversal symmetry.

\begin{acknowledgments}
This research was supported by the NRF Grant No. 2017R1D1A1B06035497 (HP) and the KIAS individual Grants No. PG013604 (HP), PG074002 (JMP) at Korea Institute for Advanced Study.
\end{acknowledgments}

\appendix

\section{Derivation of the overdamped Langevin equation}\label{sec:app_A}

We start from the formal solution of the underdamped equation, Eq.~\eqref{eq:U_Langevin}, as
\begin{equation}\label{eq:formal_sol}
    \boldsymbol{v} (t) = \frac{1}{m} \int_{-\infty}^{t} d \tau~
    e^{- \frac{1}{m} \mathbf{G} (t - \tau)}  \boldsymbol{f} \qty(\boldsymbol{x}(\tau))
    + \boldsymbol{\eta}_m (t)
\end{equation}
where
\begin{equation}
    \boldsymbol{\eta}_m (t) = \frac{1}{m} \int_{-\infty}^{t} d \tau~
    e^{- \frac{1}{m} \mathbf{G} (t - \tau)}  \boldsymbol{\xi}_T (\tau)~,
\end{equation}
with $\expval*{\boldsymbol{\xi}_T (t) \boldsymbol{\xi}_T^\rT (t')}=2\gamma T
\mathbf{I} \delta(t-t')$.

Then, the noise-noise  correlation matrix $\mathbf{Z}_m (t-t') =
\expval*{\boldsymbol{\eta}_m (t) \boldsymbol{\eta}_m^T (t')}$ can be easily obtained
as
\begin{align}\label{eq:uT}
    \mathbf{Z}_m (u) = \frac{T}{m}e^{- \frac{1}{m} \mathbf{G} u}
    ~~\textrm{and}~~\mathbf{Z}_m (-u)=\mathbf{Z}_m^\rT (u)
\end{align}
for $u>0$. Its Laplace transform becomes
\begin{align}
    \int_0^\infty du~ e^{- s u} \mathbf{Z}_m (u)
    = T \left(ms\mathbf{I} + \mathbf{G}\right)^{-1}~,
\end{align}
which becomes $T\mathbf{G}^{-1}$ in the $m\rightarrow 0$ limit. Its inverse Laplace transform
gives $\mathbf{Z}(u)=2T\mathbf{G}^{-1} \delta(u)$
and $\mathbf{Z}(-u)=\mathbf{Z}^\rT(u)$ from  Eq.~\eqref{eq:uT}.

\section{The extended FP equation for the overdamped $\xi$-dynamics}\label{sec:app_B}

We consider the $\xi$-dynamics in Eq.~\eqref{eq:U_altLangevin} and the current dynamic equation
of Eq.~\eqref{eq:current_rate}. The extended equation of motion can be rewritten as
\begin{align}\label{eqA:U_extended_Langevin}
    &\dot{\boldsymbol{q}}(t) =
    \boldsymbol{A}_q (\boldsymbol{q}(t))  + \mathbf{B}_q (\boldsymbol{q}(t))\circ \boldsymbol{\xi}_q (t) \\
\textrm{with}\quad&\boldsymbol{q}=
\begin{pmatrix}
\boldsymbol{x}\\
\Phi
\end{pmatrix}
,~\boldsymbol{A}_q(\boldsymbol{q})=
\begin{pmatrix}
\mathbf{G}^{-1} \boldsymbol{f} (\boldsymbol{x})\\
\boldsymbol{\Lambda}^\rT (\boldsymbol{x}) \mathbf{G}^{-1} \boldsymbol{f} (\boldsymbol{x})
\end{pmatrix}
,\nonumber\\
&\boldsymbol{\xi}_q=
\begin{pmatrix}
\boldsymbol{\xi}_{T_B}\\
0
\end{pmatrix}
,~
\textrm{and}\quad\mathbf{B}_q (\boldsymbol{q})=
\begin{pmatrix}
\mathbf{I} & 0\\
\boldsymbol{\Lambda}^\rT(\boldsymbol{x}) & 0
\end{pmatrix}, \label{eqA:rep1}
\end{align}
where $\dot{\boldsymbol{x}}$ in Eq.~\eqref{eq:current_rate} is replaced
in terms of $\boldsymbol{x}$ and $\boldsymbol{\xi}_{T_B}$ by Eq.~\eqref{eq:U_altLangevin}
and $\expval*{\boldsymbol{\xi}_{T_B} (t) \boldsymbol{\xi}_{T_B}^\rT (t')}
=(2 T_B/\gamma) \mathbf{I} \delta(t-t')$.

The corresponding FP operator
for the Langevin equation of Eq.~\eqref{eqA:U_extended_Langevin} with a Stratonovich-type multiplicative noise
is well established, e.g.~see the textbook~\cite{Gardiner:2010stochastic},
which yields
\begin{align}
&\hat{\mathcal{L}}_{\xi,\Phi}
    = - \boldsymbol{\partial}_q^\rT \cdot\boldsymbol{A}_q (\boldsymbol{q}) +\frac{T_B}{\gamma} \left(\boldsymbol{\partial}_q^\rT \mathbf{B}_q (\boldsymbol{q}) \right)\cdot \left(\boldsymbol{\partial}_q^\rT \mathbf{B}_q (\boldsymbol{q}) \right)^\rT,\nonumber\\&\textrm{with}\quad\boldsymbol{\partial}_q=
    \begin{pmatrix}
    \boldsymbol{\nabla}\\
    \partial_\Phi
    \end{pmatrix}~.
\end{align}
From Eq.~\eqref{eqA:rep1}, we finally obtain
\begin{equation}
\hat{\mathcal{L}}_{\xi,\Phi}=-\tilde{\boldsymbol{\nabla}}_\Phi^\rT \mathbf{G}^{-1} \boldsymbol{f} (\boldsymbol{x})+\frac{T_B}{\gamma}\tilde{\boldsymbol{\nabla}}_\Phi^\rT \cdot \tilde{\boldsymbol{\nabla}}_\Phi
\end{equation}
where  $\tilde{\boldsymbol{\nabla}}_\Phi = \boldsymbol{\nabla} + \partial_\Phi \boldsymbol{\Lambda} (\boldsymbol{x})$ is the tilted gradient
operator.

We can derive the extended FP equation with the alternative dynamic observable $\Psi$ given in
Eqs.~\eqref{def:Psi} and \eqref{eq:Phi_Psi} for the $\xi$-dynamics. This new variable changes only the
second component of $\boldsymbol{A}_q (\boldsymbol{q})$ by adding $\varphi(\boldsymbol{x})$. Therefore, we find
the extended FP operator for this case as
\begin{equation}\label{eqA:PsiFP}
\hat{\mathcal{L}}_{\xi,\Psi}=-\tilde{\boldsymbol{\nabla}}_\Psi^\rT \mathbf{G}^{-1} \boldsymbol{f} (\boldsymbol{x})+\frac{T_B}{\gamma}\tilde{\boldsymbol{\nabla}}_\Psi^\rT \cdot \tilde{\boldsymbol{\nabla}}_\Psi-\partial_\Psi \varphi (\boldsymbol{x})
\end{equation}
with  $\tilde{\boldsymbol{\nabla}}_\Psi = \boldsymbol{\nabla} + \partial_\Psi \boldsymbol{\Lambda} (\boldsymbol{x})$.

\section{The overdamped limit of the extended FP equation for the underdamped dynamics}\label{sec:app_C}

We start with the extended FP equation for the underdamped dynamics and then take
the overdamped (small-mass) limit to find the extended FP equation for the original $\eta$-dynamics.
This procedure is quite similar to that for deriving the ordinary overdamped FP equation of
Eqs.~\eqref{temp:o_FP_eq} and \eqref{temp:o_pro_current}.

The underdamped equation of motion is given by Eq.~\eqref{eq:U_Langevin}
and the accumulated current is written as
\begin{align}\label{defA:U_current}
\Phi(\Gamma)=\int_0^t dt'\boldsymbol{\Lambda}^\rT (\boldsymbol{x}(t'))\cdot\boldsymbol{v}(t')~,
\end{align}
with a trajectory $\Gamma=\{\boldsymbol{x}(t),\boldsymbol{v}(t)| t\in (0,\tau)\}$, of which the dynamic evolution
is given by
\begin{align}\label{eqA:current_rate}
\dot{\Phi}(t)=\boldsymbol{\Lambda}^\rT (\boldsymbol{x}(t))\cdot\boldsymbol{v}(t)~.
\end{align}
As $\boldsymbol{v}$ is a state variable in the underdamped description, the extended equation of motion
including $\Phi$ does not induce any multiplicative noise, in contrast to the overdamped $\xi$-dynamics.
Thus, the type of stochastic calculus such as Ito and Stratonovich is meaningless
for the FP operator.

Considering the underdamped equation of motion in Eq.~\eqref{eq:U_Langevin}
together with the above current dynamic equation, we can write
the extended FP equation  as
\begin{equation}\label{eqA:u_eFP_eq}
    \partial_t \hat{p} (\boldsymbol{x},\boldsymbol{v},\Phi,t)
    = \qty( \hat{\mathcal{L}}_{\mathrm{rev},\Phi} + \mathcal{L}_\mathrm{irr} )
    \hat{p} (\boldsymbol{x},\boldsymbol{v},\Phi,t)~,
\end{equation}
where the extended FP operator is split into the reversible and the irreversible part as
\begin{align}
    \hat{\mathcal{L}}_{\mathrm{rev},\Phi} &=
    - \tilde{\boldsymbol{\nabla}}_\Phi^\rT  \cdot \boldsymbol{v}
    - \frac{1}{m} \boldsymbol{\nabla}_v^\rT \cdot \qty(
    \boldsymbol{f}(\boldsymbol{x}) - \mathbf{B} \boldsymbol{v}) \label{eqA:erev}\\
    \mathcal{L}_\mathrm{irr} &=
    \frac{\gamma}{m} \boldsymbol{\nabla}_v^\rT \cdot
    \qty ( \boldsymbol{v} + \frac{T}{m} \boldsymbol{\nabla}_v ) \label{eqA:eirr}
\end{align}
with $\tilde{\boldsymbol{\nabla}}_\Phi = \boldsymbol{\nabla} + \partial_\Phi \boldsymbol{\Lambda} (\boldsymbol{x})$ , the velocity gradient
operator $\boldsymbol{\nabla}_v$, and the magnetic-field matrix
$\mathbf{B}=\mathbf{G}_\textrm{a}$ (the antisymmetric part of $\mathbf{G}$).
For convenience, we include the Lorentz term ($\mathbf{B}\boldsymbol{v}$) in the reversible operator.

Note that the information of the current variable $\Phi$ is fully contained
in the tilted position gradient operator $\tilde{\boldsymbol{\nabla}}_\Phi$ in the exactly same way
 as in the overdamped case (Appendix~\ref{sec:app_B}).
 Hence, we can easily expect that  the overdamped (small-mass) limit of the extended FP equation
 should be the same as that for the ordinary FP equation of  Eqs.~\eqref{temp:o_FP_eq} and \eqref{temp:o_pro_current}, except for replacing $\boldsymbol\nabla$ by $\tilde{\boldsymbol{\nabla}}_\Phi$,
 which is presented in Eq.~\eqref{eq:exo_FP_eq_eta2}.

For completeness, we briefly sketch the standard procedure for the
small-mass expansion~\cite{Risken:1996FokkerPlanck, Chun2018emergence, Durang2015overdamped}.
The strategy is to find a series expansion of the distribution $\hat{p}(\boldsymbol{x},\boldsymbol{v}, \Phi,t)$ in terms of the orthonormal basis of the irreversible operator $\mathcal{L}_\mathrm{irr}$.
Then, the extended FP equation provides a hierarchy of coupled differential equations for the expansion
coefficients known as the Brinkman's hirarchy~\cite{Brinkman1956brownian, Risken:1996FokkerPlanck}.
In the small-mass limit, most of higher-order coefficients can be neglected in this expansion and
the overdamped extended FP equation can be rather easily obtained.

First, it is convenient to rotate the operator ${\mathcal{L}}_\mathrm{irr}$  in a Hermitian form by introducing
\begin{align}
\bar{\mathcal{L}}_\mathrm{irr} = (\sqrt{p^\mathrm{ss}(\boldsymbol{v})})^{-1}
\mathcal{L}_\mathrm{irr} \sqrt{p^\mathrm{ss}(\boldsymbol{v})}
\end{align}
with
\begin{equation}
    p^\mathrm{ss}(\boldsymbol{v}) =
    \frac{m}{2 \pi T}
    \exp [- \frac{m}{2 T} \qty|\boldsymbol{v}|^2]~,
\end{equation}
which is the steady-state solution  of $\mathcal{L}_\mathrm{irr}$,
satisfying $\mathcal{L}_\mathrm{irr} p^\mathrm{ss}(\boldsymbol{v})=0$.

In terms of the classical analogue of the bosonic ladder operators
\begin{equation}
    \boldsymbol{a}_\pm = \frac{1}{2} \sqrt{\frac{m}{T}} \boldsymbol{v}
    \mp \sqrt{\frac{T}{m}} \boldsymbol{\nabla}_v~,
\end{equation}
the rotated operator is written as
\begin{equation}\label{defA:L_irr_bar_a}
    \bar{\mathcal{L}}_\mathrm{irr}
    = - \frac{\gamma}{m} \boldsymbol{a}_+^\rT \cdot \boldsymbol{a}_-~,
\end{equation}
which is reminiscent of the Hamiltonian of the quantum harmonic oscillator.
Its orthonormal eigenfunctions $\bar\psi_{n_1,n_2} (\boldsymbol{v})$ are well known with
two quantum numbers $(n_1,n_2)$, which are simply related to each other
by the raising and lowering ladder operators, $\boldsymbol{a}_\pm$.
The ground-state eigenfunction is given as $\bar\psi_{0,0} (\boldsymbol{v})=\sqrt{p^\mathrm{ss}(\boldsymbol{v})}$.
It is useful to consider the same rotation on the reversible operator $\hat{\mathcal{L}}_{\mathrm{rev},\Phi}$
as
\begin{align}
    \bar{\mathcal{L}}_{\mathrm{rev},\Phi}
    = &-\sqrt{\frac{T}{m}} \tilde{\boldsymbol{\nabla}}_\Phi^\rT \cdot \boldsymbol{a}_-  - \sqrt{\frac{T}{m}} \left(\tilde{\boldsymbol{\nabla}}_\Phi-\frac{\boldsymbol{f}(\boldsymbol{x})}{T}\right)^\rT\cdot  \boldsymbol{a}_+ \nonumber\\
    &\quad\quad - \frac{1}{m} \boldsymbol{a}_+^\rT \mathbf{B}~  \boldsymbol{a}_-~.\label{defA:L_rev_bar_a}
\end{align}

The FP equation~\eqref{eqA:u_eFP_eq} can be rewritten in terms of the rotated operators as
\begin{equation}\label{eqA:u_FP_eq_rot}
    \partial_t \bar{p} (\boldsymbol{x},\boldsymbol{v},\Phi,t)
    = \qty( \bar{\mathcal{L}}_{\mathrm{rev},\Phi} + \bar{\mathcal{L}}_\mathrm{irr} )
    \bar{p} (\boldsymbol{x},\boldsymbol{v},\Phi,t)~,
\end{equation}
with the transformed distribution
\begin{align}\label{eqA:barp}
\bar{p} (\boldsymbol{x},\boldsymbol{v},\Phi,t)\equiv (\bar{\psi}_{0,0} (\boldsymbol{v}))^{-1} ~\hat{p} (\boldsymbol{x},\boldsymbol{v},\Phi,t)~.
\end{align}
We expand
$\bar{p}(\boldsymbol{x},\boldsymbol{v},\Phi,t)$ in terms of the orthonormal eigenfunctions
as
\begin{equation}\label{eqA:u_ex_p}
    \bar{p}(\boldsymbol{x},\boldsymbol{v},\Phi,t) =
    \sum_{n_1, n_2=0}^\infty  \hat{C}_{n_1, n_2} \bar{\psi}_{n_1,n_2} (\boldsymbol{v})
\end{equation}
with the expansion coefficient $\hat{C}_{n_1, n_2} = \hat{C}_{n_1, n_2} (\boldsymbol{x},\Phi,t)$.
By putting this expansion form into Eq.~\eqref{eqA:u_FP_eq_rot} and focussing on
the lower-order terms for small $m$~\cite{Chun2018emergence}, we find
\begin{align}\label{eqA:Brink1}
    &\partial_t \hat{C}_{0,0}  =
    -\sqrt{\frac{T}{m}} \tilde{\boldsymbol{\nabla}}_\Phi^\rT \cdot \hat{\boldsymbol{C}}_1~, \\
    &\mathbf{G} \hat{\boldsymbol{C}}_1  =
    - \sqrt{mT} \left(\tilde{\boldsymbol{\nabla}}_\Phi-\frac{\boldsymbol{f}(\boldsymbol{x})}{T}\right)
    \hat{C}_{0,0} + \mathcal{O}(m) \label{eqA:Brink2}\\
 &\textrm{with}\quad\hat{\boldsymbol{C}}_1=
\begin{pmatrix}
\hat{C}_{1,0}\\
\hat{C}_{0,1}
\end{pmatrix}\nonumber~.
\end{align}

The overdamped extended distribution $\hat{p}_\eta (\boldsymbol{x},\Phi,t)$ is obtained by integrating out
the velocity variable of $\hat{p}(\boldsymbol{x},\boldsymbol{v},\Phi,t)$. Using Eqs.~\eqref{eqA:barp} and
\eqref{eqA:u_ex_p} with the orthogonality property of
the eigenfunctions, we find  $\hat{C}_{0,0}=\hat{p}_\eta (\boldsymbol{x},\Phi,t)$.
Combining Eqs.~\eqref{eqA:Brink1} and \eqref{eqA:Brink2},
we find the overdamped extended FP operator
as
\begin{align}
\hat{\mathcal{L}}_{\eta,\Phi}
    = -  \tilde{\boldsymbol{\nabla}}_\Phi^\rT \mathbf{G}^{-1} \left(\boldsymbol{f} (\boldsymbol{x})
    - T\tilde{\boldsymbol{\nabla}}_\Phi\right)~.
\end{align}

\section{Stochastic thermodynamics for the
$\eta$-dynamics}\label{sec:app_D}

In this appendix, we develop the stochastic thermodynamics
for the $\eta$-dynamics. As the noise $\boldsymbol{\eta}_T$ has
a singular character in the noise-noise correlation,
the framework of the stochastic thermodynamics should be
reexamined. Thus,
we start from the underdamped dynamics
where the stochastic thermodynamics is well established.

\subsection{FP operator and probability distribution}
The underdamped FP equation with the Lorentz force is
\begin{equation}\label{eqA:u_FP_eq}
    \partial_t p (\boldsymbol{x},\boldsymbol{v},t)
    = \qty( \mathcal{L}_\mathrm{rev} + \mathcal{L}_\mathrm{irr} )
    p (\boldsymbol{x},\boldsymbol{v},t)~,
\end{equation}
where the ordinary FP operators $\mathcal{L}_\mathrm{rev}$ and $\mathcal{L}_\mathrm{irr}$
are given in Eqs.~\eqref{eqA:erev} and \eqref{eqA:eirr} with the replacement $\tilde{\boldsymbol{\nabla}}_\Phi$ by $\boldsymbol{\nabla}$.
With the same series expansion, we finally get
the same equations as  Eqs.~\eqref{eqA:Brink1} and \eqref{eqA:Brink2} for the expansion coefficients
$C_{n_1,n_2}(\boldsymbol{x},t) = \int d\Phi \hat{C}_{n_1,n_2} (\boldsymbol{x},\Phi,t)$, again by replacing $\tilde{\boldsymbol{\nabla}}_\Phi$ with $\boldsymbol{\nabla}$.
As we have
\begin{align}
p_\eta(\boldsymbol{x},t)=\int d\boldsymbol{v} ~p (\boldsymbol{x},\boldsymbol{v},t) =C_{0,0}
\end{align}
and
\begin{align}
\boldsymbol{J}_\eta(\boldsymbol{x},t)=\sqrt{\frac{T}{m}}~\boldsymbol{C}_1 = \mathbf{G}^{-1} \left(\boldsymbol{f} (\boldsymbol{x}) - T\boldsymbol{\nabla}\right)p_\eta(\boldsymbol{x},t)~,\label{eqA:J_eta}
\end{align}
then the overdamped FP operator becomes
\begin{align}
\mathcal{L}_\eta
    &= -  \boldsymbol{\nabla}^\rT \mathbf{G}^{-1} \left(\boldsymbol{f} (\boldsymbol{x})
    - T\boldsymbol{\nabla}\right)~.
\end{align}

The underdamped distribution for the small-mass limit is obtained from the
expansion like in Eq.~\eqref{eqA:u_ex_p}
\begin{align}\label{eqA:approx_dis}
p (\boldsymbol{x},\boldsymbol{v},t)&\approx \bar\psi_{0,0}(\boldsymbol{v}) \left[  C_{0,0}\bar\psi_{0,0}(\boldsymbol{v}) +
\boldsymbol{C}_1^\rT \cdot
\begin{pmatrix}
\bar\psi_{1,0}(\boldsymbol{v})\\
\bar\psi_{0,1}(\boldsymbol{v})
\end{pmatrix}
 + \cdots\right]\nonumber\\
&\approx p^\mathrm{ss}(\boldsymbol{v})\left[p_\eta(\boldsymbol{x},t)+\frac{m}{T} \boldsymbol{J}_\eta^\rT(\boldsymbol{x},t)\cdot \boldsymbol{v}+\cdots\right]\\
&\approx p_\eta(\boldsymbol{x},t)\left(\frac{m}{2 \pi T}\right)
    \exp [- \frac{m}{2 T} \left|\boldsymbol{v}-\frac{\boldsymbol{J}_\eta(\boldsymbol{x},t)}{p_\eta(\boldsymbol{x},t)}\right|^2]~,\nonumber
\end{align}
where we used the eigenfunction relations as $\bar\psi_{1,0} (\boldsymbol{v})=v_1 \sqrt{\frac{m}{T}}\bar{\psi}_{0,0} (\boldsymbol{v})$ and
$\bar\psi_{0,1} (\boldsymbol{v})=v_2 \sqrt{\frac{m}{T}}\bar{\psi}_{0,0} (\boldsymbol{v})$.
As expected, the average local velocity
$\langle \boldsymbol{v}\rangle_v=\int d\boldsymbol{v}~\boldsymbol{v} p (\boldsymbol{x},\boldsymbol{v},t)= \boldsymbol{J}_\eta(\boldsymbol{x},t)$.
This approximate probability distribution form near the overdamped limit is also valid for the normal $\xi$-dynamics.

\subsection{average current and entropy production}

The average current  of Eq.~\eqref{eqA:current_rate} is calculated as
\begin{align}\label{eqA:ave_current}
\langle \dot{\Phi}\rangle &= \int d\boldsymbol{x} d\boldsymbol{v}~ \boldsymbol{\Lambda}^\rT(\boldsymbol{x})\cdot \boldsymbol{v}~ p (\boldsymbol{x},\boldsymbol{v},t)\nonumber\\
&= \int d\boldsymbol{x}~ \boldsymbol{\Lambda}^\rT(\boldsymbol{x}) \cdot \boldsymbol{J}_\eta(\boldsymbol{x},t)~,
\end{align}
where we used $\boldsymbol{J}_\eta(\boldsymbol{x},t)=\int d\boldsymbol{v}~\boldsymbol{v} p (\boldsymbol{x},\boldsymbol{v},t)$.
This form is the same as that for the normal overdamped $\xi$-dynamics.
However, note that
the probability current $\boldsymbol{J}_\eta (\boldsymbol{x},t) \neq \boldsymbol{J}_\xi (\boldsymbol{x},t)$.

The total entropy production rate is given as~\cite{Spinney2012entropy, Kwon2016unconventional}
\begin{align}
\langle \dot{S}^\textrm{tot}\rangle = \frac{m^2}{\gamma T}
\int d\boldsymbol{x} d\boldsymbol{v}~  \frac{|\boldsymbol{J}_\textrm{irr}(\boldsymbol{x},\boldsymbol{v},t)|^2}{p(\boldsymbol{x},\boldsymbol{v},t)}~,
\end{align}
where the irreversible probability current is given by
\begin{equation}
    \boldsymbol{J}_\mathrm{irr} (\boldsymbol{x}, \boldsymbol{v}, t)
    = -\frac{\gamma}{m} \qty (\boldsymbol{v}
    + \frac{T}{m}
    \boldsymbol{\nabla}_v)
    p(\boldsymbol{x}, \boldsymbol{v}, t)~.
\end{equation}
Note that the Lorenz term is not included for the irreversible current, so the total
entropy production here does not involve the so-called unconventional entropy production term
and becomes the Clausius entropy in the steady state~\cite{Kwon2016unconventional,Park2018entropy,Chun2018microscopic}.

Using the approximate distribution function in Eq.~\eqref{eqA:approx_dis} for the overdamped limit,
we find
\begin{equation}
    \boldsymbol{J}_\mathrm{irr} (\boldsymbol{x}, \boldsymbol{v}, t)
    \approx  -\frac{\gamma}{m} \frac{\boldsymbol{J}_\eta(\boldsymbol{x},t)}{p_\eta(\boldsymbol{x},t)}   p(\boldsymbol{x}, \boldsymbol{v}, t)~,
\end{equation}
which yields the entropy production rate in the overdamped limit as
\begin{align}\label{eqA:EP_total_zeta_1}
\langle \dot{S}^\textrm{tot}\rangle
\approx \frac{\gamma}{T}
\int d\boldsymbol{x}~ \frac{|\boldsymbol{J}_\eta(\boldsymbol{x},t)|^2}{p_\eta(\boldsymbol{x},t)}~.
\end{align}
This form is also standard for the normal overdamped Langevin system with the probability current $\boldsymbol{J}_\eta (\boldsymbol{x},t)$.

\subsection{stochastic currents}

The stochastic work current done by the external force $\boldsymbol{f}_\textrm{ext} (\boldsymbol{x})$ is given as
\begin{align}\label{eqA:W_rate1}
\dot{W}(t)= \boldsymbol{f}_\textrm{ext}^\rT(\boldsymbol{x}(t))\cdot\boldsymbol{v}(t)~,
\end{align}
in the underdamped dynamics. Thus, it is natural to define its overdamped $\eta$-dynamics as
\begin{align}\label{eqA:W_rate2}
\dot{W}(t)= \boldsymbol{f}_\textrm{ext}^\rT(\boldsymbol{x}(t))\circ\dot{\boldsymbol{x}}(t)~,
\end{align}
which is consistent with its average in Eq.~\eqref{eqA:ave_current}.

The stochastic heat current dissipated into the reservoir is given as
\begin{align}\label{eqA:Q_rate1}
\dot{Q}(t) &= \qty(\gamma \boldsymbol{v}(t) -\boldsymbol{\xi}_T(t))^\rT\circ \boldsymbol{v}(t)~,\\
&=\left(\boldsymbol{f}(\boldsymbol{x}(t))-\mathbf{B}\boldsymbol{v}(t)\right)^\rT\cdot\boldsymbol{v}(t)-\frac{d}{dt} \left(\frac{1}{2}m\qty|\boldsymbol{v}(t)|^2\right)~,
\end{align}
where the underdamped equation of motion, Eq.~\eqref{eq:U_Langevin}, is used.
The Lorentz term vanishes due to the antisymmetric property of $\mathbf{B}$.
The above form manifests that the heat current is not of the generic type in Eq.~\eqref{eqA:current_rate}.
However, in the overdamped $\eta$-dynamics by taking the small-mass limit, the heat current becomes the generic current type
of Eq.~\eqref{eq:current_rate} as
\begin{align}\label{eqA:Q_rate2}
\dot{Q}(t)= \boldsymbol{f}^\rT(\boldsymbol{x}(t))\circ\dot{\boldsymbol{x}}(t)~.
\end{align}

The stochastic entropy production rate in the overdamped $\eta$-dynamics can be obtained from
the system Shannon entropy production rate and the Clausius entropy production due to the heat dissipation.
Taking the time derivative of the stochastic system Shannon entropy, we find
\begin{align}\label{eq:sto_sysEP_eta}
\dot{S}^\textrm{sys}_\eta (t) &= -\frac{d}{dt} \ln p_\eta (t) =-\frac{\partial_t p_\eta (t)}{p_\eta (t)} -\frac{{\boldsymbol{\nabla}}^\rT p_\eta (t)}{p_\eta (t)}\circ \dot{\boldsymbol{x}} (t)\nonumber\\
&=-\frac{\partial_t p_\eta (t)}{p_\eta (t)}-\frac{\dot{Q} (t)}{T} +\frac{(\mathbf{G} \boldsymbol{J}_\eta (\boldsymbol{x}(t),t))^\rT}{Tp_\eta (t)}\circ \dot{\boldsymbol{x}} (t)
\end{align}
with $p_\eta (t) = p_\eta (\boldsymbol{x}(t),t)$. We used Eq.~\eqref{temp:o_pro_current} for the final equality, which yields
\begin{align}\label{eqA:EP_sto}
\dot{S}^\textrm{tot}_\eta (t) =\dot{S}^\textrm{sys}_\eta (t) +\frac{\dot{Q}(t)}{T}
=-\frac{\partial_t p_\eta (t)}{p_\eta (t)}+
\boldsymbol{\Lambda}_S^\rT (\boldsymbol{x}(t),t)
\circ \dot{\boldsymbol{x}} (t)
\end{align}
with $\boldsymbol{\Lambda}_S (\boldsymbol{x},t)=\mathbf{G} \boldsymbol{J}_\eta (\boldsymbol{x},t)/(Tp_\eta (\boldsymbol{x},t))$.
We note that this expression is not of the generic type for currents in Eq.~\eqref{eq:current_rate} due to the presence of the
first term in the right hand side of the equation and the explicit time-dependency of the weight function. But, in the calculation of its average, the first term does not contribute and Eq.~\eqref{eqA:EP_total_zeta_1}
is easily recovered using Eq.~\eqref{eqA:ave_current}. Furthermore, in the steady state, the first term
vanishes by definition and the weight function is time-independent as in Eq.~\eqref{eq:EP0}, thus the stochastic total entropy production
rate becomes the generic type.

\subsection{entropy production for the $\xi$-dynamics}

It is worthwhile to derive explicitly the stochastic entropy production rate and its average in the naive $\xi$-dynamics.
At the first glance, one may think that the result may be trivial because the $\xi$-dynamics is a standard overdamped dynamics
with a white Gaussian noise. It turns out to be rather subtle due to the mismatch of the asymmetric friction tensor
$\mathbf{G}$ and the symmetric diffusion matrix $(T_B/\gamma)\mathbf{I}$.
When we take the distribution equivalence
$p_\xi (t) = p_\xi (\boldsymbol{x} (t), t) = p_\eta (\boldsymbol{x} (t), t)$ into account,
it is obvious that the time derivative of the stochastic system entropy has the same form as in Eq.~\eqref{eq:sto_sysEP_eta}.
Following the similar procedure as above, we get
\begin{align}\label{eqA:EP_sto_xi}
\dot{S}^\textrm{tot}_\xi (t) =\dot{S}^\textrm{sys}_\xi (t) +\frac{\dot{Q}(t)}{T}
=-\frac{\partial_t p_\xi (t)}{p_\xi (t)}+
\boldsymbol{\Lambda}^\rT_S (\boldsymbol{x}(t),t) \circ \dot{\boldsymbol{x}} (t)~,
\end{align}
where the heat dissipation rate $\dot{Q} (t)$ is defined in Eq.~\eqref{eqA:Q_rate2}
and $\boldsymbol{\Lambda}_S (\boldsymbol{x},t)=\mathbf{G} \boldsymbol{\mathcal{J}}_\xi (\boldsymbol{x},t)/(Tp_\xi (\boldsymbol{x},t))$
with the effective probability current $\boldsymbol{\mathcal{J}}_\xi (\boldsymbol{x},t)$ defined in Eq.~\eqref{eq:eff_cur2}.
From the relation between currents ($\boldsymbol{\mathcal{J}}_\xi (\boldsymbol{x},t)=\boldsymbol{{J}}_\eta (\boldsymbol{x},t)$) in Eq.~\eqref{eq:JJ_rel}, we find that the stochastic entropy expressions
for both dynamics are identical, i.e.~$\dot{S}^\textrm{tot}_\eta (t) =\dot{S}^\textrm{tot}_\xi (t) \equiv\dot{S}^\textrm{tot} (t)$.

However, its average should be different as expected from Eq.~\eqref{eq:Phi_diff}. In the $\xi$-dynamics,
we find
\begin{align}
\langle \dot{S}^\textrm{tot}\rangle_\xi &=
\int d\boldsymbol{x}~ \frac{(\mathbf{G} \boldsymbol{\mathcal{J}}_\xi (\boldsymbol{x},t))^\rT}{Tp_\xi (\boldsymbol{x},t)}\cdot \boldsymbol{{J}}_\xi (\boldsymbol{x},t)\nonumber\\
&=\frac{\gamma}{T}\int d\boldsymbol{x}~ \frac{|\boldsymbol{\mathcal{J}}_\xi(\boldsymbol{x},t)|^2}{p_\xi(\boldsymbol{x},t)}
+\langle {\boldsymbol{\nabla}}^\rT \mathbf{G}_\textrm{a}^{-1}\boldsymbol{f}\rangle_\xi\nonumber\\
&=
\langle \dot{\Sigma}\rangle_\xi -\langle\varphi_S\rangle_\xi ~\label{eqA:EP_Diff}
\end{align}
where the scalar function $\varphi_S(\boldsymbol{x},t)=-T \boldsymbol{\nabla}^\rT \mathbf{G}_\mathrm{a}^{-1} \boldsymbol{\Lambda}_S (\boldsymbol{x},t)$ with $\boldsymbol{\Lambda}_S(\boldsymbol{x},t)=\mathbf{G} \boldsymbol{\mathcal{J}}_\xi(\boldsymbol{x},t)/(Tp_\xi(\boldsymbol{x},t))$ from Eq.~\eqref{eq:phi}
and the corresponding alternative dynamic observable $\Sigma$, defined as $\dot{\Sigma} (t)= \dot{S}^\textrm{tot}(t) +\varphi_S (\boldsymbol{x}(t),t)$,
satisfies $\langle \dot{\Sigma}\rangle_\xi=\langle \dot{S}^\textrm{tot}\rangle_\eta$, consistent with
Eq.~\eqref{eq:Phi_diff}.

\section{Modified TUR with an arbitray initial state}\label{sec:app_E}

We consider a time-dependent perturbation on the force as
\begin{equation}\label{defA:pert_force}
    \boldsymbol{f}_{\theta} (\boldsymbol{x},t) = \boldsymbol{f} (\boldsymbol{x})
    + \theta \mathbf{G}
    \frac{\boldsymbol{\mathcal{J}}_\xi (\boldsymbol{x},\bar{t})}
    {p_\xi(\boldsymbol{x}.\bar{t})}~, \quad \textrm{with}\quad \bar{t}=(1+\theta)t~,
\end{equation}
where the perturbation term is defined by the quantities for the unperturbed dynamics at a later time
scaled by a factor of $1+\theta$ with respect to the perturbed dynamics at time $t$.

Similar to the steady-state case, we can easily show that the perturbed distribution function at $t$
is identical to the unperturbed one at the scaled time $\bar{t}$, i.e.~$p_{\xi,\theta}(\boldsymbol{x},t) = p_\xi(\boldsymbol{x},\bar{t})$.
This can be easily checked as
the perturbed probability current with this relation
\begin{align}
    \boldsymbol{J}_{\xi,\theta}(\boldsymbol{x},t)
    &= \qty( \mathbf{G}^{-1}  \boldsymbol{f}_\theta (\boldsymbol{x})
    - \frac{T_B}{\gamma} \boldsymbol{\nabla} )
    p_{\xi,\theta}(\boldsymbol{x},t)\\
    &=\boldsymbol{J}_\xi(\boldsymbol{x},\bar{t}) + \theta \boldsymbol{\mathcal{J}}_\xi (\boldsymbol{x},\bar{t})
\end{align}
satisfies the Fokker-Planck equation for the perturbed dynamics
\begin{align}
\partial_t p_{\xi,\theta}(\boldsymbol{x},t)=-\boldsymbol{\nabla}^\rT \cdot \boldsymbol{J}_{\xi,\theta}(\boldsymbol{x},t)~,
\end{align}
where we used the relation $\boldsymbol{\nabla}^\rT \cdot \boldsymbol{\mathcal{J}}_\xi(\boldsymbol{x},t)=\boldsymbol{\nabla}^\rT\cdot \boldsymbol{J}_\xi(\boldsymbol{x},t)$.

It is simple to show that
\begin{equation}
    \boldsymbol{\mathcal{J}}_{\xi,\theta}(\boldsymbol{x},t)
    = \mathbf{G}^{-1}  \qty( \boldsymbol{f}_\theta (\boldsymbol{x})
    - T \boldsymbol{\nabla} ) p_{\xi,\theta} (\boldsymbol{x},t)
    = ( 1 + \theta ) \boldsymbol{\mathcal{J}}_\xi (\boldsymbol{x},\bar{t})~,
\end{equation}
which yields
\begin{align}
\expval*{{\Psi}}_{\xi,\theta} &=\int_0^t dt'd\boldsymbol{x}~\boldsymbol{\Lambda}^\rT(\boldsymbol{x})\cdot \boldsymbol{\mathcal{J}}_{\xi,\theta}(\boldsymbol{x},t')\nonumber\\
&=\int_0^{(1+\theta)t} d\bar{t}'d\boldsymbol{x}~\boldsymbol{\Lambda}^\rT(\boldsymbol{x})\cdot \boldsymbol{\mathcal{J}}_\xi(\boldsymbol{x},\bar{t}')\nonumber\\
&=\int_0^{(1+\theta)t} d\bar{t}'\expval*{\dot{\Psi}(\bar{t}')}_\xi~.
\end{align}
Thus, we get
\begin{align}
\lim_{\theta\rightarrow 0}\partial_\theta \expval*{{\Psi}}_{\xi,\theta}= t \expval*{\dot{\Psi}({t})}_\xi
=t \expval*{\dot{\Phi}({t})}_\eta~.
\end{align}

Using the expression for the trajectory probability in Eq.~\eqref{eq:path_prob}
with the same initial condition for any value of $\theta$,
we can also find the Fisher information from Eq.~\eqref{eq:Fisher_info} as
\begin{align}
    I(\theta) = \frac{1}{2(1+\theta)}
    \qty(
    \frac{\gamma}{T_B}\int_0^{(1+\theta)t} dt'
     d\boldsymbol{x}
    \frac{
    | \boldsymbol{\mathcal{J}}_\xi (\boldsymbol{x},t') |^2
    }{p_\xi (\boldsymbol{x},t')})~.
\end{align}
Setting $\theta=0$ and using Eq.~\eqref{eq:JJ_rel} , we find
\begin{align}
    I(0) = \frac{\gamma}{2T_B}
    \int_0^{t} dt'
     d\boldsymbol{x}
    \frac{
    | \boldsymbol{J}_\eta (\boldsymbol{x},t') |^2
    }{p_\eta (\boldsymbol{x},t')}=\frac{T}{2T_B} \langle\Delta S^\textrm{tot}\rangle_\eta~,
\end{align}
where $\langle\Delta S^\textrm{tot}\rangle_\eta$ is the total entropy production for  duration time $t$
and Eq.~\eqref{eqA:EP_total_zeta_1} is used.

Finally, we have a modified TUR for the overdamped $\eta$-dynamics with an arbitrary initial state for a finite
duration time $t$ as
\begin{align}\label{eqA:TUR_arb}
\frac{\mathrm{Var}_\eta \left[ \Phi  \right]}
    {\expval*{t\dot{\Phi}(t)}_\eta^2}
    \langle\Delta S^\textrm{tot}\rangle_\eta
    \ge 2k_\textrm{B} \frac{T_B}{T}~,
\end{align}
where the Boltzmann constant $k_\textrm{B}$ is restored.

\section{Extension to higher dimensions}\label{sec:app_F}

Consider a three-dimensional dynamics with a magnetic field in one direction.
The two dimensional plane perpendicular to the magnetic field is spanned by
the first and second component of the position vector $\boldsymbol{x}$.
Its third component describes the magnetic field direction.
Then, the friction coefficient tensor in Eq.~\eqref{eq:Gamma} is given by
\begin{align}\label{eqA:Gamma}
    \mathbf{G} =
    \begin{pmatrix}
        \gamma & - B & 0\\
        B & \gamma & 0 \\
        0 & 0 & \gamma
    \end{pmatrix}~.
\end{align}

The derivation procedure of the modified TUR is basically the same as before in the two-dimensional case, so
we list here only the differences. The symmetric part of the correlation matrix in Eq.~\eqref{eq:Zs} becomes
\begin{align}\label{eqA:Zs}
    \mathbf{Z}_\mathrm{s}  (u)= \frac{2T_B}{\gamma}
    \begin{pmatrix}
        1 & 0& 0\\
        0& 1 & 0 \\
        0 & 0 & \frac{T}{T_B}
    \end{pmatrix}\delta(u) \equiv 2\mathbf{D}_\mathrm{s} \delta(u)~,
\end{align}
and the naive probability current in Eq.~\eqref{temp:o_pro_n_current} becomes
\begin{equation}\label{eqA:o_pro_n_current}
    \boldsymbol{J}_\xi (\boldsymbol{x},t) = \qty(\mathbf{G}^{-1}
     \boldsymbol{f}(\boldsymbol{x}) -\mathbf{D}_\mathrm{s}\boldsymbol{\nabla} )
    p_\xi(\boldsymbol{x},t)~.
\end{equation}
For the $\xi$-dynamics, $T_B/\gamma$ should be replaced by $\mathbf{D}_\mathrm{s}$ everywhere.

The key replacement is in the trajectory probability in Eq.~\eqref{eq:path_prob},
which yields the Fisher information as
\begin{align}
    I(\theta) &= \frac{t}{2}
    \int d\boldsymbol{x}
    \frac{
    \qty(\boldsymbol{\mathcal{J}}_\xi^\mathrm{ss} (\boldsymbol{x}))^\rT \mathbf{D}_\mathrm{s}^{-1} \boldsymbol{\mathcal{J}}_\xi^\mathrm{ss} (\boldsymbol{x})
    }{p_\xi^\mathrm{ss} (\boldsymbol{x})}\\
    &<
    \frac{t\gamma}{2T_B}
    \int d\boldsymbol{x}
    \frac{
    | \boldsymbol{\mathcal{J}}_\xi^\mathrm{ss} (\boldsymbol{x}) |^2
    }{p_\xi^\mathrm{ss} (\boldsymbol{x})}~,\label{eqA:inequa}
\end{align}
where the third component of $| \boldsymbol{\mathcal{J}}_\xi^\mathrm{ss} (\boldsymbol{x}) |^2$
makes the inequality for $T_B<T$. As the entropy production for the $\eta$-dynamics
in Eq.~\eqref{eq:EP_total_zeta_1}
does not change in form, we finally get the same modified TUR in Eq.~\eqref{eq:main_result}
for three dimensions. The extensions to higher dimensions and also for arbitrary initial states are obvious.
But, due to the inequality in Eq.~\eqref{eqA:inequa}, the TUR equality cannot be attained for nonzero $B$
in three or higher dimensions.

\section{Exact solutions in the solvable model}\label{sec:app_G}

Consider the FP equation in Eqs.~\eqref{temp:o_FP_eq} and \eqref{temp:o_pro_current}
for the $\eta$-dynamics with the linear force $\boldsymbol{f}(\boldsymbol{x}) = - \mathbf{F} \boldsymbol{x}$
in Eq.~\eqref{eq:Fx}, given as
\begin{align}
 \partial_t p(\boldsymbol{x},t)
    = \boldsymbol{\nabla}^\rT \cdot \qty( \mathbf{A}  \boldsymbol{x}
    + \mathbf{D}  \boldsymbol{\nabla} ) p(\boldsymbol{x},t)~,
\end{align}
with the drift matrix $\mathbf{A} = \mathbf{G}^{-1}\mathbf{F}$ and the asymmetric diffusion matrix
$\mathbf{D} = T \mathbf{G}^{-1}$. We dropped the subscript $\eta$ for brevity.
It is easy to show that the covariant matrix $\mathbf{C}\equiv \langle \boldsymbol{x}\boldsymbol{x}^\rT\rangle=\mathbf{C}^\rT$ should satisfy
\begin{align}
\mathbf{A}\mathbf{C}+\mathbf{C}\mathbf{A}^\rT=\mathbf{D}+\mathbf{D}^\rT~,
\end{align}
in the steady state~\cite{Gardiner:2010stochastic,Risken:1996FokkerPlanck,Lee2020exactly}.
Then, we find
\begin{align}\label{eqA:Cov}
\mathbf{C}=\frac{\gamma T}{\gamma k +\epsilon B} \mathbf{I}=\frac{T}{k(1+\epsilon_0 B_0)} \mathbf{I}~,
\end{align}
where the dimensionless parameters are used with $\epsilon_0=\epsilon/k$ and $B_0=B/\gamma$.

We consider an accumulated current $\Phi$ of the generic type in Eq.~\eqref{def:neq_current} with $\boldsymbol{\Lambda}(\boldsymbol{x})=\mathbf{J} \boldsymbol{x}$. The linearity in $\boldsymbol{x}$
makes the exact calculation of their means and variances in the steady state rather simpler.
The extended FP equation in Eqs.~\eqref{eq:exo_FP_eq_eta1}
and \eqref{eq:exo_FP_eq_eta2} is written as
\begin{equation}
    \partial_t \hat{p}(\boldsymbol{x},\Phi,t)
    = \tilde{\boldsymbol{\nabla}}_\Phi^\rT \cdot \qty( \mathbf{A}  \boldsymbol{x}
    + \mathbf{D}  \tilde{\boldsymbol{\nabla}}_\Phi ) \hat{p}(\boldsymbol{x},\Phi,t)~,
\end{equation}
with the tilted gradient operator
$\tilde{\boldsymbol{\nabla}}_\Phi = \boldsymbol{\nabla} + \partial_\Phi (\mathbf{J} \boldsymbol{x})$.
By multiplying a function of variables and integrating out over
$\boldsymbol{x}$ and $\Phi$ on both sides of the extended FP equation, we can derive the dynamic equation
of its average in terms of the averages of other functions.
As we are interested in the steady-state average only, we take
the initial condition $\hat{p}(\boldsymbol{x},\Phi,0)=p^\textrm{ss} (\boldsymbol{x})\delta(\Phi)$.

For the first moment of $\Phi$, we can easily obtain, through simple integrations by parts,
\begin{equation}
    \partial_t \expval*{\Phi }
    = \Tr{\mathbf{J}^\rT(\mathbf{D}-\mathbf{A}\mathbf{C})}
    \equiv \phi~,
\end{equation}
where $\Tr{\mathbf{X}}$ stands for the
trace of a matrix $\mathbf{X}$.
As $\phi$ is a constant of time, we get the average accumulated current as $\expval*{\Phi }=t \phi $.
Of course, we can get the same result by integrating Eq.~\eqref{eqA:ave_current}.

It is straightforward but rather involved to calculate $\expval*{\Phi^2 }$.
Here, we only sketch the calculation procedure.
First, we obtain the dynamic equation for the second moment of $\Phi$ as
\begin{equation}\label{eq:second_moment}
    \partial_t \expval*{\Phi^2}
    = 2 \Tr{\phi \mathbf{D}^\rT \mathbf{J} t
    - \mathbf{J}^\rT \mathbf{A} \mathbf{M} (t)
    + \mathbf{J}^\rT \mathbf{D} \mathbf{J} \mathbf{C}}
\end{equation}
with $\mathbf{M}(t) =
\expval*{\Phi \boldsymbol{x}\boldsymbol{x}^\rT}$.
This equation contains higher moments $\mathbf{M}(t)$ and
its dynamic equation will again contain higher moments, but
only 4th moments of the position variable $\boldsymbol{x}$.
Since $p^\textrm{ss}(\boldsymbol{x})$ is Gaussian as in Eq.~\eqref{eq:ss_Gaussian},
the standard Wick's theorem splits  a 4th moment into a combination of second moments,
which can be given by the covariant matrix $\mathbf{C}$.
Finally, the dynamic equation for $\mathbf{M} (t)$ is given by the following closed equation,
\begin{align}
    \partial_t \mathbf{M} (t)
    =& - \mathbf{A} \mathbf{M}^\rT (t)
    - \mathbf{M} (t) \mathbf{A}^\rT
    + \phi (\mathbf{C} + (\mathbf{D}+\mathbf{D}^\rT)t) \nonumber\\
    &+ \mathbf{A} \mathbf{C} \mathbf{J}\mathbf{C}
    + \mathbf{C} \mathbf{J}^\rT \mathbf{C} \mathbf{A}^\rT~.
\end{align}
This is a linear differential equation for $\mathbf{M}(t)$,
so that the formal solution is given by
\begin{equation}
    \mathbf{M} (t) = \phi \mathbf{C} t
    + \mathbf{H} - \mathbf{E} (t) \mathbf{H} \mathbf{E}^\rT (t)~,
\end{equation}
where an auxiliary symmetric matrix $\mathbf{H}$
is defined by the relation
\begin{equation}\label{eq:mat_rel_H}
    \mathbf{A} \mathbf{C} \mathbf{J}\mathbf{C}
    + \mathbf{C} \mathbf{J}^\rT \mathbf{C} \mathbf{A}^\rT =
    \mathbf{A} \mathbf{H} + \mathbf{H} \mathbf{A}^\rT~,
\end{equation}
and a matrix $\mathrm{E} (t)$ is defined by
\begin{equation}
    \mathbf{E} (t) =
    e^{-\mathbf{A} t} =
    e^{- \frac{\gamma k + \epsilon B}{B^2 + \gamma^2} t}
    \begin{pmatrix}
        \cos (\omega t) & - \sin (\omega t)\\
        \sin (\omega t) & \cos (\omega t)
    \end{pmatrix}
\end{equation}
with $\omega = (k B-\gamma \epsilon)/(B^2 + \gamma^2)$.
By inserting the solution into Eq.~\eqref{eq:second_moment}
and using the relation $\partial_t \mathrm{Var} [\Phi]
= \partial_t \expval*{\Phi^2} - 2 \phi^2 t$,
we obtain
\begin{equation}
    \partial_t \mathrm{Var} [\Phi]
    = 2 D_\Phi
    + 2 \Tr{\mathbf{J}^\rT \mathbf{A}
    \mathbf{E}(t) \mathbf{H} \mathbf{E}^\rT (t)}~,
\end{equation}
where the diffusion coefficient $D_\Phi$ is
given by
\begin{equation}
    D_\Phi =
    \Tr{\mathbf{J}^\rT (
    \mathbf{D} \mathbf{J} \mathbf{C}
    - \mathbf{A} \mathbf{H})}~.
\end{equation}
By introducing another auxiliary symmetric matrix $\mathbf{K}$, for convenience, such that
\begin{equation}\label{eq:mat_rel_K}
    \mathbf{J}^\rT \mathbf{A} + \mathbf{A}^\rT \mathbf{J}
    = \mathbf{K} \mathbf{A} + \mathbf{A}^\rT \mathbf{K}~,
\end{equation}
we finally reach the formal solution
\begin{equation}
    \mathrm{Var} [\Phi]
    = 2 D_\Phi t
    + \Tr{ \mathbf{K} \qty( \mathbf{H}- \mathbf{E}(t) \mathbf{H} \mathbf{E}^\rT (t) ) }~.
\end{equation}

The formal solution allows us to obtain the variance
by solving the linear equations in the elements of
$\mathbf{H}$ and $\mathbf{K}$
provided by Eq.~\eqref{eq:mat_rel_H} and \eqref{eq:mat_rel_K}.
By determining the auxiliary matrices for
$\mathbf{J} = \mathbf{W}$, $\mathbf{Q}$, and $\mathbf{S}$,
we obtain the variance for work, heat, and entropy production.

\bibliography{paper}

\end{document}